\documentclass[fleqn,usenatbib]{mnras}

\usepackage{newtxtext,newtxmath}

\usepackage[T1]{fontenc}

\DeclareRobustCommand{\VAN}[3]{#2}
\let\VANthebibliography\thebibliography
\def\thebibliography{\DeclareRobustCommand{\VAN}[3]{##3}\VANthebibliography}

\usepackage{graphicx}	
\usepackage{amsmath}	
\usepackage[export]{adjustbox}
\usepackage{pdflscape}
\usepackage{xcolor, color, soul}
\usepackage[most]{tcolorbox}
\usepackage[normalem]{ulem}

\title[Frequencies of CB discs]{On the frequencies of circumbinary discs in protostellar systems}

\author[D. Elsender et al.]{
Daniel Elsender,$^{1}$ \thanks{E-mail: de296@exeter.ac.uk (DE)}
Matthew R. Bate,$^{1}$
Ben S. Lakeland,$^{1}$
Eric L. N. Jensen$^{2}$
and Stephen H. Lubow$^{3}$
\\
$^{1}$Department of Physics and Astronomy, University of Exeter, Stocker Road, Exeter EX4 4QL, UK\\
$^{2}$Department of Physics and Astronomy, Swarthmore College, Swarthmore, PA 19081, USA\\
$^{3}$Space Telescope Science Institute, 3700 San Martin Drive, Baltimore, MD 21218, USA}

\date{Accepted XXX. Received YYY; in original form ZZZ}

\pubyear{2023}

\begin{document}
\label{firstpage}
\pagerange{\pageref{firstpage}--\pageref{lastpage}}
\maketitle

\begin{abstract}
We report the analysis of circumbinary discs formed in a radiation hydrodynamical simulation of star cluster formation. We consider both pure binary stars and pairs within triple and quadruple systems. The protostellar systems are all young (ages < $10^5$ yrs).  We find that the systems that host a circumbinary disc have a median separation of $\approx 11$ au, and the median  characteristic radius of the discs is $\approx 64$ au. We find that $89$ per cent of pure binaries with semi-major axes $a<1$ au have a circumbinary disc, and the occurrence rate of circumbinary discs is bimodal with log-separation in pure binaries with a second peak at $a \approx 50$ au. Systems with $a>100$ au almost never have a circumbinary disc. The median size of a circumbinary disc is between $\approx 5-6\ a$ depending on the order of the system, with higher order systems having larger discs relative to binary separation.  We find the underlying distributions of mutual inclination between circumbinary discs and binary orbits from the simulation are in good agreement with those of observed circumbinary discs.
\end{abstract}
\begin{keywords}
accretion, accretion discs -- binaries: general -- hydrodynamics -- methods: numerical -- planets and satellites: formation -- protoplanetary discs 
\end{keywords}



\section{Introduction}

Stars commonly form in binary systems \citep[e.g.][]{abt_levy_1976_multiplicity_1976ApJS...30..273A,duquennoy_mayor_multiplicity_1991A&A...248..485D,ghez1993multiplicity,fisher_turbulent_2004,raghavan_multiplicity_2010ApJS..190....1R}. It is widely accepted that most stellar binaries form due to fragmentation of the protostellar core, or protostellar discs \citep[e.g.][]{boss_fragmentation_1979,boss_protostellar_1986,bonnell_formation_1994,bate_sphng_1995MNRAS.277..362B,kratter_fragmentation_2006,clarke_pseudo-viscous_2009}. When forming, binaries are not able to form directly with a separation of less than $a\sim 10$ au due to the opacity limit of fragmentation \citep[][]{boss_protostellar_1986,bate_collapse_1998}. Close binaries ($a<10$ au) are likely to have migrated inwards due to accretion, interactions with discs, and/or dynamical interactions with other stars \citep[][]{artymowicz_orbital_evolution_1991,bate_formation_2002,tokovinin_formation_2020}. During the star formation process a binary system may capture gas with high angular momentum from its local environment to form a natal circumbinary disc around a binary. The dynamic interactions between the disc and the binary can dominate the evolution of the natal disc \citep[][]{papaloizou_pringle_tidal_torques_1977MNRAS.181..441P,lin_papaloizou_tidal_torques_1979MNRAS.186..799L,lin_papaloizou_structure_of_cb_discs_1979MNRAS.188..191L}. The outcome of such interactions are dependent on properties of the binaries themselves, for example the semi-major axis of the system ($a$) and the mass ratio of the component stars ($q$) \citep[see][for example]{artymowicz_lubow_binary_disc_interaction_1994ApJ...421..651A}. Binaries with a semi-major axis $a < 50$ au are observed to truncate the outer radius of circumstellar discs \citep[][]{jensen_separation_1996ApJ...458..312J} whilst close binaries ($a\lesssim10$ au) are likely to  have circumbinary (CB) discs \citep[][]{harris_taurus_multiple_systems_2012ApJ...751..115H}. Such discs are also known as discs in the "P"-type configuration \citep[][]{dvorak_orbit_types_1982OAWMN.191..423D}.


CB discs were first identified in the 1990s, beginning with a combination of resolved CB discs around multiple systems: GG Tauri \citep{kawabe_discovery_1993,dutrey1994images}, UZ Tauri E \citep[][]{jensen_high-resolution_1996,mathieu_uz_1996}, UY Aurigae \citep{duvert_disks_1998}; and inferred discs around known multiple systems: GW Orionis \citep[][]{mathieu_submillimeter_1995}, AK Sco and V4046 Sgr \citep{jensen_evidence_1997}, DQ Tau \citep{mathieu_classical_1997}. Over the past 30 years some of the complexities of these systems have been revealed, such as disc size, and system orientation. The discs around GW Ori have been a focus of interest recently due to the geometry of the system \citep[][]{czekala_architecture_2017,bi_gw_2020,kraus_triple-star_2020} with evidence that the disc surrounding this system is inclined $\sim 45 \degr$ to the stellar orbital plane, and \citet{kraus_triple-star_2020} suggest this misalignment is due to disc warping and tearing. In addition to this some observed features in CB discs have been attributed to disc warps. \citet{marino_shadows_2015} discovered a warped inner disc of HD 142527 casting shadows on the outer disc. Warps have previously been proposed to be the cause of non-axisymmetric features of CB discs, as in the case of the T-Tauri disc TW Hya \citep[][]{rosenfeld_disk-based_2012}. Like discs around single objects, CB discs can extend out to 100s of au, for example the CB disc of V892 Tau has CO gas emission detected at 200 au from the central binary \citep{long_architecture_2021}. CB discs can live into the Class II phase, with an approximate lifetime of 2 Myr \citep{kraus_role_2012}. However, whilst more than forty CB discs have now been detected \citep[see][and references therein]{czekala_coplanar_2021} their detections are relatively scarce compared to discs about single objects. For example, \citet{akeson_binaries_in_taurus_2019ApJ...872..158A} detected no new CB discs even in their survey of multiple systems in the Taurus-Auriga star-forming region with a non-detection limit that corresponds to a gas mass of approximately 0.04 Jupiter masses.

It is not known at what rate CB discs should occur, or their typical lifetimes, due to difficulties observing them. On scales $\lesssim 100$ au optically thick dust emission can make it difficult to detect companions separated $\lesssim 50$ au \citep[][]{tobin_vla_2016}. Despite these difficulties there have been some observations of very young CB discs in embedded Class 0 objects \citep[e.g.][]{harris_alma_2018,sadavoy_dust_2018,hsieh_vla1623a_2020ApJ...894...23H,belinski_orbital_2022}. \citet{tobin_vlaalma_2018} find circum-multiple dust emission around eight of the nine Class 0 multiple objects they observed in the Perseus region, although they find none around Class I multiple objects. This may be an indication of the typical lifetime of such discs.

Due to the turbulent nature of molecular clouds, chaotic accretion can take place \citep[][]{bate_cluster_formation_2003MNRAS.339..577B,mckee_theory_2007} and cause misalignments between the binary orbital plane and circumbinary disc \citep[][]{bate_chaotic_star_formation_2010MNRAS.401.1505B,bate_2018_10.1093/mnras/sty169}. There are observations of inclined circumbinary discs, for example KH 15D \citep[][]{chiang_circumbinary_2004} and IRS 43 \citep[][]{brinch_misaligned_2016}. Recently there has been additional theoretical work done on the evolution of circumbinary disc inclinations. Notably, initially misaligned discs around eccentric binaries can evolve to a polar alignment \citep[][]{martin_lubow_polar_alignment_i_2017ApJ...835L..28M}. Such a disc has been confirmed by observations: the CB disc around the system HD 98800 \citep[][]{kennedy_circumbinary_2019}. Misaligned discs can also occur due to Kozai-Lidov oscillations \citep[][]{von_zeipel_sur_1910,kozai_secular_1962,lidov_evolution_1962}, in which disc inclination and eccentricity are exchanged \citep[][]{martin_kozai-lidov_2014, fu_kozai-lidov_2015,aly_efficient_2020}. During these oscillations dust that is sufficiently coupled to the gas undergoes oscillations on the same timescale as the gas. The dust may have a different distribution to that of the gas due to a higher radial velocity drift \citep[][]{zagaria_dust_2021}, and during periods of high disc eccentricity the dust may break into rings \citep[][]{martin_eccentric_2022}. \citet{kuruwita_role_2019} suggest that molecular cloud cores that are more turbulent may produce more massive CB discs, potentially extending their lifetimes. This may have an impact on planet formation, if the lifetime of the disc is shorter than the timescale of alignment, and planet formation is sufficiently quick, then the planet may be left misaligned to the binary.

Even though CB disc detections are somewhat uncommon, there is a growing catalogue of circumbinary planets in particular around eclipsing binaries, where the planet transits the binary objects. The eclipsing binary method is an effective way to detect close-in planets around binary objects due to the high likelihood that a planet will transit the objects \citep[see][]{martin_circumbinary_2015}. Most of these have been detected by the \textit{Kepler} spacecraft \citep[e.g.][]{Doyle1602,welsh_kepler_2012Natur.481..475W}, with the \textit{Transiting Exoplanet Survey Satellite} (\textit{TESS}) recently discovering two \citep[][]{kostov_TESS_TOI_1338_2020AJ....159..253K,kostov_TIC_172900988_2021AJ....162..234K}. This indicates that CB discs not only form but can live long enough for planets to grow. Using eclipsing binaries to detect planets mostly relies on the planets being coplanar with the binary orbit. So there is a selection effect. However with the recent detection of a circumbinary planet using radial velocities, which are less restricted to an edge-on configuration, this may change \citep[][]{standing_first_2023}. If coplanarity is preferred for close binaries then these detections suggest that binaries may have a planet occurrence rate close to that of single stars; if the distribution of planetary orbit inclination is more uniform, then the occurrence rate could be higher than that of single stars \citep[][]{armstrong_abundance_2014}. Observing transits of misaligned CB planets becomes a more difficult problem due to the geometry of the system, however there have been efforts to develop a method to detect such planets \citep[see][for example]{martin_planets_2014}.

Given that discs are a natural consequence of star formation, and most stars form in pairs, it is important to understand the statistical properties of CB discs. In addition to this, since there is a growing catalogue of planets orbiting a central binary, the properties of the protoplanetary circumbinary discs also impact planet formation. Knowing how often we expect a disc to form around a binary, how long they tend to live for, and how they are inclined relative to their binary has significant implications for planet formation.

In this paper we present: our method of characterising these discs from star cluster formation simulations in Section \ref{sec:method}, the statistics of the circumbinary discs in Section \ref{sec:disc stats}, a comparison of simulated disc statistics with observed disc statistics in Section \ref{sec:discussion}, and our conclusions in Section \ref{sec:conc}.

\section{Methods}
\label{sec:method}

The discs studied in this paper are from a calculation originally published by \citet{bate_metallicity_2019MNRAS.484.2341B}. For a full description of the methods used to perform the calculation and the properties of the star cluster the reader is directed there. Here we give only a brief overview of the method used to perform the calculations in Section \ref{sec:rad hydro calc}, the properties of the cloud and cluster in Section \ref{sec:cloud props}, and the algorithm used to extract the circumbinary discs and their properties in Section \ref{sec:cb disc char}

\subsection{The radiation hydrodynamical calculation}
\label{sec:rad hydro calc}

The star cluster formation calculation was performed using the smoothed particle hydrodynamics (SPH) code, \textsc{sphNG}, developed from the original version of \citet{benz_review_1990nmns.work..269B} and \citet{benz_1990ApJ...348..647B}, but significantly adapted  \citep[e.g.][]{bate_sphng_1995MNRAS.277..362B,price_2007MNRAS.374.1347P,whitehouse_2005MNRAS.364.1367W,whitehouse_2006MNRAS.367...32W} and parallelised using both \textsc{OpenMP} and \textsc{MPI}.

The calculation employs the combined radiative transfer and diffuse interstellar medium (ISM) model of \citet{bate_keto_10.1093/mnras/stv451}.  The diffuse ISM model is similar to that of \citet{glover_metal_poor_gas_2012MNRAS.426..377G} but using a simplified chemical model.  It is particularly important for thermal modelling at low densities (e.g., hydrogen number densities $n_{\rm H} \lesssim 10^5$ cm$^{-3}$). The radiative transfer employs the flux-limited diffusion radiative transfer method of \citet{whitehouse_2005MNRAS.364.1367W} and \citet{whitehouse_2006MNRAS.367...32W}.

The gas has an ideal gas equation of state for pressure $p=\rho T_{\text{gas}}\mathcal{R}/\mu$, where $\rho$ is the gas density, $T_\text{gas}$ is the gas temperature, $\mathcal{R}$ is the gas constant, and $\mu$ is the mean molecular weight of gas. The hydrogen and helium mass fractions are $X = 0.70$ and $Y=0.28$, respectively, setting $\mu = 2.38$.


In the calculation, protostellar collapse was followed through its first core phase and into the second collapse phase \citep[][]{larson_1969MNRAS.145..271L}. As the density increases, the time-step required decreases; due to this sink particles \citep[][]{bate_sphng_1995MNRAS.277..362B} were inserted when the density exceeded $10^{-5}~\text{g cm}^{-3}$, just before stellar core formation at $\sim 10^{-3}~\text{g cm}^{-3}$.

We note that the calculation does not include numerous physical effects. There are no magnetic fields, protostellar outflows and radiation from inside the sink particle accretion radius (0.5 au) is neglected. The calculation also begins from idealised initial conditions. Nevertheless, previous studies have found this calculation, and others with similar physics included, to reproduce observed statistics of stellar and circumstellar disc properties \citep{bate_metallicity_2014MNRAS.442..285B,bate_2018_10.1093/mnras/sty169,bate_metallicity_2019MNRAS.484.2341B,elsender_metallicity_2021}.


\subsection{Properties of the cloud and cluster}
\label{sec:cloud props}

The calculation uses an identical density and velocity structure to \citet{bate_2012_10.1111/j.1365-2966.2011.19955.x} and \citet{bate_metallicity_2014MNRAS.442..285B}, with a full description in \citet{bate_2012_10.1111/j.1365-2966.2011.19955.x}. The initial conditions consist of a spherical, uniform-density molecular cloud with $500~ \text{M}_\odot$ of gas and a radius of $0.404\text{pc}$. This yields an initial density of $1.2 \times 10^{-19} \text{g cm}^{-3}$, a number density of hydrogen of $n_\text{H} \approx 6 \times 10^4 ~ \text{cm}^{-3}$, and a free fall time of $t_\text{ff}=6 \times 10^{12} ~\text{s}$ ($1.9 \times 10^{5} \text{yr}$). The cloud is given an initial turbulent velocity field \citep[][]{Ostriker_turbulence_2001ApJ...546..980O,bate_cluster_formation_2003MNRAS.339..577B}, generated from a divergence free random Gaussian velocity field with a power spectrum $P(k) \propto k^{-4}$, where $k$ is the wavenumber on a $128^3$ uniform grid with SPH particle velocities interpolated from this. The velocity field is normalised such that the kinetic energy from the turbulence is equal to the magnitude of the gravitational potential energy of the cloud (the r.m.s. Mach number is $\mathcal{M} = 13.7$ at $10$~K).


The calculation used $3.5 \times 10^7$ SPH particles to model the cloud, providing sufficient resolution to resolve the local Jeans mass throughout the calculation, necessary to model fragmentation of the collapsing molecular cloud \citep[][]{bate_resolution_1997MNRAS.288.1060B,truelove_resolution_1997ApJ...489L.179T,whitworth_jeans_instab_1998MNRAS.296..442W,boss_jeans_2000ApJ...528..325B,hubber_resolution_2006A&A...450..881H}.

The radiation hydrodynamical calculation was evolved to $t = 1.2~t_\mathrm{ff}$ $(228\ 300\ \mathrm{yr})$. By this time the calculation had produced 255 stars or brown dwarfs. The first sink particle was created in the calculation at around $1.15\times 10^5$ yr ($0.606 ~t_\mathrm{ff}$), which is about halfway into the total run time of the calculation. The total mass of gas converted into stars and brown dwarfs was $90.1\ \mathrm{M}_\odot$, from the initial cloud mass of $500\ \mathrm{M}_\odot$, giving a mean mass of the stars and brown dwarfs of $0.35 \pm 0.04 \mathrm{M}_\odot$. At the end of the calculation there were 26, 13 and 9 binary, triple, and quadruple systems, respectively. 

In the calculation there were $14$ stellar mergers. They occur when two sink particles pass within $6\ \mathrm{R}_\odot$ ($0.03$ au) of each other. Although binary orbits with periastron distances down to this separation can be modelled, the accretion radius of a sink particle is $0.5$ au, and so only discs larger than a few au are resolved. This inhibits the formation of very close systems because of the absence of gas on very small scales.

\subsection{Disc characterisation}
\label{sec:cb disc char}

In characterising the properties of the CB discs, we follow the method of \citet{bate_2018_10.1093/mnras/sty169}.  The discs formed in these calculation continually evolve due to a range of hydrodynamical processes \citep{bate_2018_10.1093/mnras/sty169}, so to study the properties of the discs we extract properties many times during the calculation.  The idea is that if observers examined a very large sample of real protostellar systems the ensemble would consist of systems with many different ages. We take snapshots of the discs every $0.0025t_\text{ff}$ (476 yr) for each protostar. In the solar metallicity calculation this gives $16048$ instances of circumstellar discs, and $1515$ instances of circumbinary discs.

Previous similar studies of discs \citep{bate_2018_10.1093/mnras/sty169,elsender_metallicity_2021} have found that the basic disc properties (masses and sizes) are in quite good agreement with the observed discs of young (Class 0) protostellar systems. Those studies considered discs in single and multiple systems, however they did not study circumbinary discs in detail. They did include CB discs along with discs of triple and quadruple systems (circum-system discs) but CB discs were not studied separately.  In this paper we examine CB discs in much greater detail. 

\subsubsection{Circumstellar discs}

Prior to discussing how we characterise circumbinary discs, first it is essential to discuss how we characterise circumstellar discs as discs can form about either one, or both of the protostars in the binary pair. The discs around individual components are usually referred to as circumstellar discs. First let us consider a single protostar. We sort the SPH gas and sink particles by distance from the sink particle representing the protostar. Starting with the nearest particle, we say this is part of the protostellar disc if:
\begin{enumerate}
    \item it has not already been assigned to another disc,
    \item the instantaneous ballistic orbit of the particle has an apastron distance of less than 2000 au,
    \item its orbital eccentricity $e < 0.3$.
\end{enumerate}
If these criteria are met the mass of the particle is added to the system mass, and the position and velocity of the centre of mass of the system are calculated. This is repeated for the next closest particle. Particles are only considered if they are closer than 2000 au. If a sink particle is found when traversing the list of particles the identity of the sink is recorded, and the process of adding mass to the circumstellar disc is ended. In other words, the disc does not include any gas particles more distant than the closest protostar. 

Whilst the above algorithm gives reasonable results for the properties of discs it sometimes finds discs with very large radii and low mass; when visualising these `discs' it is clear that they are just made up of small amounts of rotating infalling material. This gas is part of the envelope about the protostar. To exclude these `discs' we ignore anything labelled as a `disc' with mass $< 0.03 \mathrm{M}_\odot$ for which $>63$ per cent of this mass is contained within a radius of $>300$ au. In addition to this we exclude `discs' with a radius containing $63$ per cent of the disc mass that is more than three times greater than the radius containing $50$ per cent of the disc mass. These cuts have been used in previous studies \citep[][]{bate_2018_10.1093/mnras/sty169,elsender_metallicity_2021} and are effective at separating discs from envelope material.

Following \citet{bate_2018_10.1093/mnras/sty169}, when referring to the radius of a disc we mean the characteristic radius, $r_\mathrm{c}$, from the truncated power-law density profile \citep[][]{fedele_rings_and_gaps_in_ppd_2017A&A...600A..72F,2017A&A...606A..88T}
\begin{equation}
\label{eq:radial power law}
    \Sigma(r) = \Sigma_\mathrm{c} \left(\frac{r}{r_\mathrm{c}}\right) ^ {-\gamma} \exp{\left[-\left(\frac{r}{r_\mathrm{c}}\right) ^ {\left(2-\gamma\right)}\right]},
\end{equation}
where $\gamma$ is the power-law radial density profile of the disc, and $\Sigma_\mathrm{c}/e$ is the gas surface density at $r_\mathrm{c}$. When $\gamma<2$, $r_\mathrm{c}$ is always the radius that contains $(1-1/e)$ of the total disc mass ($\sim 63.2$ per cent; \citet{bate_2018_10.1093/mnras/sty169}). Equation \ref{eq:radial power law} only gives a sensible distribution for $\gamma<2$. Therefore, we do not fit Equation \ref{eq:radial power law} to a disc. Rather, we obtain $r_\mathrm{c}$ simply by measuring the radius that contains $63.2$ per cent of the total disc mass.

\subsubsection{Circumbinary discs}

Many of the discs formed in these calculation are found to be in bound multiple systems. The scope of this paper is to discuss the discs surrounding bound `pairs' of protostars. Sometimes there are binary systems within a higher order system (e.g. a triple system consisting of a binary and a companion). We will be clear whether the binary is a pure binary or a binary within a higher order system when discussing results. 

The mass of CB discs is computed in a similar way to that of circumstellar discs. However, instead of searching radially outward from a sink particle we calculate the centre of mass of the binary pair and search radially from there.  Note, the sink particles of the bound pair do not cause the radial search to stop in this case. Similarly to the circumstellar discs we list all particles within $2000$ au of the centre of mass of the system in order of distance. When going through the list of particles we ignore particles that are contained in any of the component discs (i.e. circumprimary or circumsecondary). Applying the same criteria as above, we obtain a small number of `CB discs' that are in fact envelope material. To avoid this we added one more requirement to extract CB discs. We discard any `CB discs' that contain less than $50$ per cent of the total system disc mass (i.e. sum of circumbinary, primary, secondary disc masses). This value was chosen as it means each CB disc will have a significant mass relative to circumstellar discs, and was found to be effective at ignoring envelope material whilst keeping what would usually be considered to be CB discs. This percentage of total mass may seem to be high; this is discussed more in the next section.

Upon incorporating these criteria the number of instances of CB discs is reduced to $1465$, from $1515$. There are $650$, $409$, and $406$ instances of CB discs in pure binaries, triples, and quadruples respectively.

\begin{figure}
    \centering
    \includegraphics[width=0.98\linewidth]{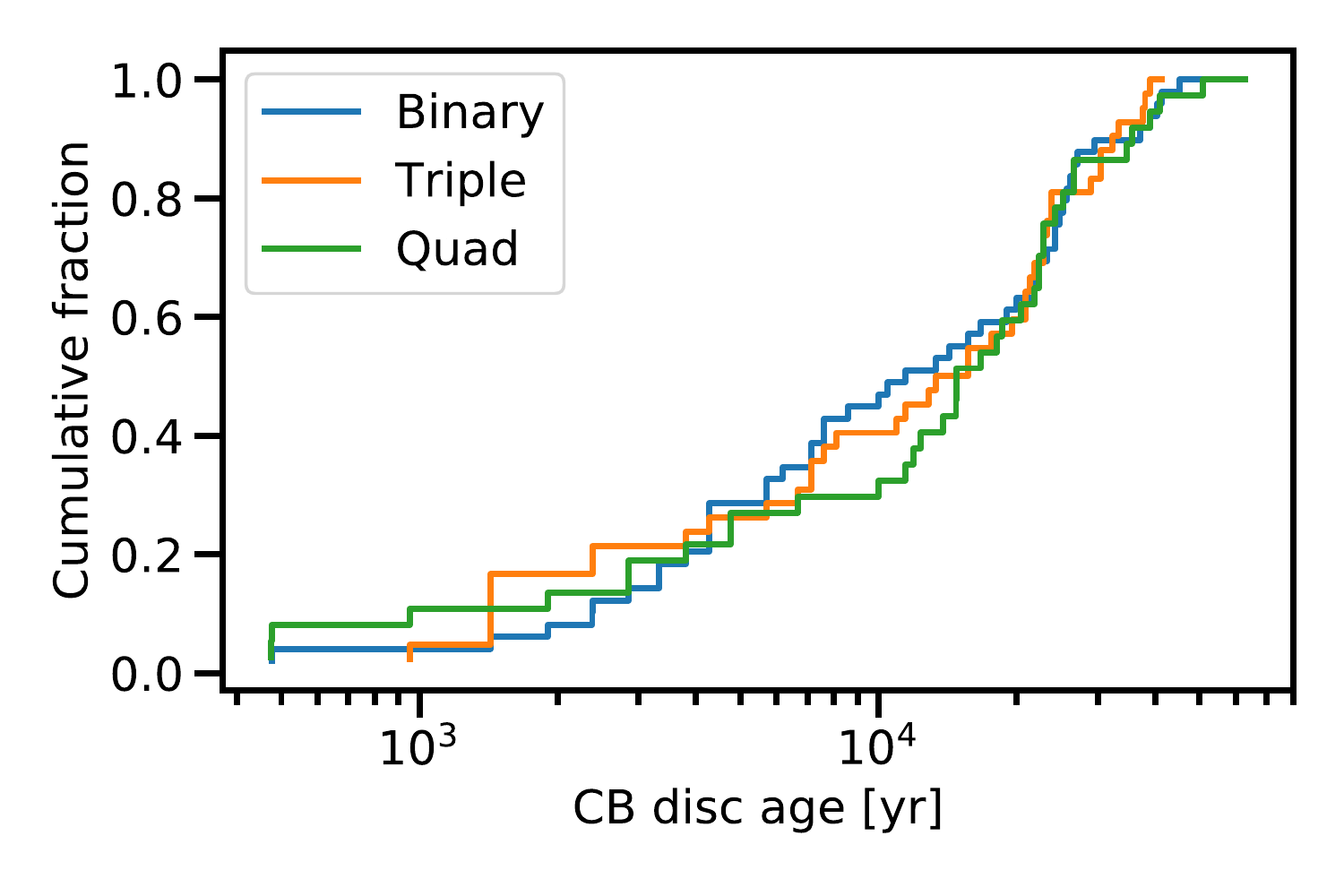} \vspace{-12pt}
    \caption{Cumulative distributions of the ages of discs taken at either the end of the calculation or at the final snapshot they are recorded in, i.e. the disc is no longer classified or the central object is no longer a binary (e.g., a merger may take place, the binary may become unbound, or another components may exchange into the original binary). Most of the discs in all multiple systems live longer than $10^4$ yr. There is no significant multiplicity dependence on the age distributions of the CB discs.}
    \label{fig:cb_disc_lifetime}
\end{figure}

\section{Results}
\label{sec:disc stats}

In this section we present statistics on CB discs found within the solar metallicity star formation calculation of \citet{bate_metallicity_2019MNRAS.484.2341B}. \citet{elsender_metallicity_2021} studied the disc properties of all four calculations from \citet{bate_metallicity_2019MNRAS.484.2341B} and found some dependence on metallicity. Typically disc size decreases with metallicity, and discs and orbits of bound pairs tend to be less aligned at lower metallicity. Therefore we expect there to be some difference in the statistics of CB discs in calculations differing metallicity (e.g. smaller disc size, and a lower occurrence rate of CB discs at lower metallicity), however this is outside the scope of this study.

We summarise the ages of CB discs in the solar-metallicity calculation in Section \ref{sec:lifetime of discs} and the occurrence rate of CB discs in different systems in Section \ref{sec:freq of discs}. In Section \ref{sec: acc disc stats} we present the numerous different statistics of the CB discs in the cluster simulations, e.g. radius/mass distributions, disc-orbit mutual inclinations, radius in units of binary separation, disc mass to system mass ratio. The statistics generated from the calculation are from snapshots taken throughout the simulation, every $476\ \mathrm{yr}$. This means that if a disc is present in the calculation for multiple snapshots then the properties of this disc will be included that number of times. Because the discs are continually evolving, the properties of the disc change enough between snapshots that no two snapshots are the same. This does mean, however, that not all instances are strictly independent from each other.

\subsection{Ages of discs}
\label{sec:lifetime of discs}

As the calculation is only run for a set time period ($1.20 ~t_\mathrm{ff}$) and the first stars form at $0.606 ~t_\mathrm{ff}$, the oldest protostars can only be $\approx 10^5$ yr old at most. When considering the frequency of CB discs, it would be useful to know how old the discs are. If discs tend to be short lived it would make them less likely to be observed. In Fig. \ref{fig:cb_disc_lifetime} we plot the cumulative age distribution of the CB discs at the point where they fail to meet the criteria laid out in Section \ref{sec:cb disc char} for CB discs in binary, triple, and quadruple systems. The distributions in Fig. \ref{fig:cb_disc_lifetime} suggest that it does not matter if the disc is around a bound pair in a binary or a hierarchical system; they tend to live, regardless of the system, at least up until ages of $\approx 10^4$ yr. We note that during the simulation discs come and go, i.e. a disc may form around a binary and then cease to be present at a later time. Additionally, sometimes discs may briefly fail to meet the criteria laid out in the previous section and then go on to satisfy them at a later point. No CB discs found in the simulation have ages greater than $8 \times 10^4$ years. The older discs tended to be the most massive ones to form in the simulations (0.1-1 $\mathrm{M}_\odot$) These systems are very young compared to those that would typically be observed, however there are some cases of CB discs about comparatively young objects. For example, young CB discs have been observed in L1551 NE \citep[][]{takakuwa_L1551_NE_2012ApJ...754...52T}, L1448 IRS3B \citep[][]{tobin_triple_2016}, IRAS 16293-2422 A \citep[][]{maureira_orbital_2020}, and VLA1623 A \citep[][]{hsieh_vla1623a_2020ApJ...894...23H}. This is an indication that CB discs form quickly about protostellar binaries. In the case of L1448 IRS3B, the system is estimated to be less than 150,000 years old. 

During their evolution, CB discs can be disrupted in many ways due to the highly dynamic and chaotic process of star formation that can lead to them to having their properties altered or being destroyed entirely \citep[see][which discusses the many processes that can drive disc evolution]{bate_2018_10.1093/mnras/sty169}. We see this on many occasions in this simulation. For example, the CB disc formed around sink particles 3 and 6 (numbered in the order of formation in the calculation) grows in mass until it reaches $\approx 0.4 \mathrm{M}_\odot$, after which its mass suddenly decreases by three orders of magnitude. The sudden drop in disc mass is caused by the fragmentation of the CB disc to form a protostar, forming a triple system. CB discs that are misaligned and undergo fragmentation may result in triple systems that are also misaligned. The disc around sinks 9 and 13 is the most massive disc in the simulation ($\approx 0.8 \mathrm{M}_\odot$), rapidly growing towards the end of the calculation, until there is a significant increase in dynamical interactions and the disc is completely destroyed. Another example of how CB discs can have their properties altered is the disc around sinks 52 and 176. This disc's mass goes through periodic increases and decreases in mass. This however is not due to periodic busts of accretion, but rather the capture of a third outer star that then orbits the system sweeping through the disc, accreting material away from it but not totally destroying it. These and other events can be viewed in the disc mosaic animation that was published as supplementary material with \citet{bate_metallicity_2019MNRAS.484.2341B}.

\begin{figure}
    \centering
    \includegraphics[width=0.98\linewidth]{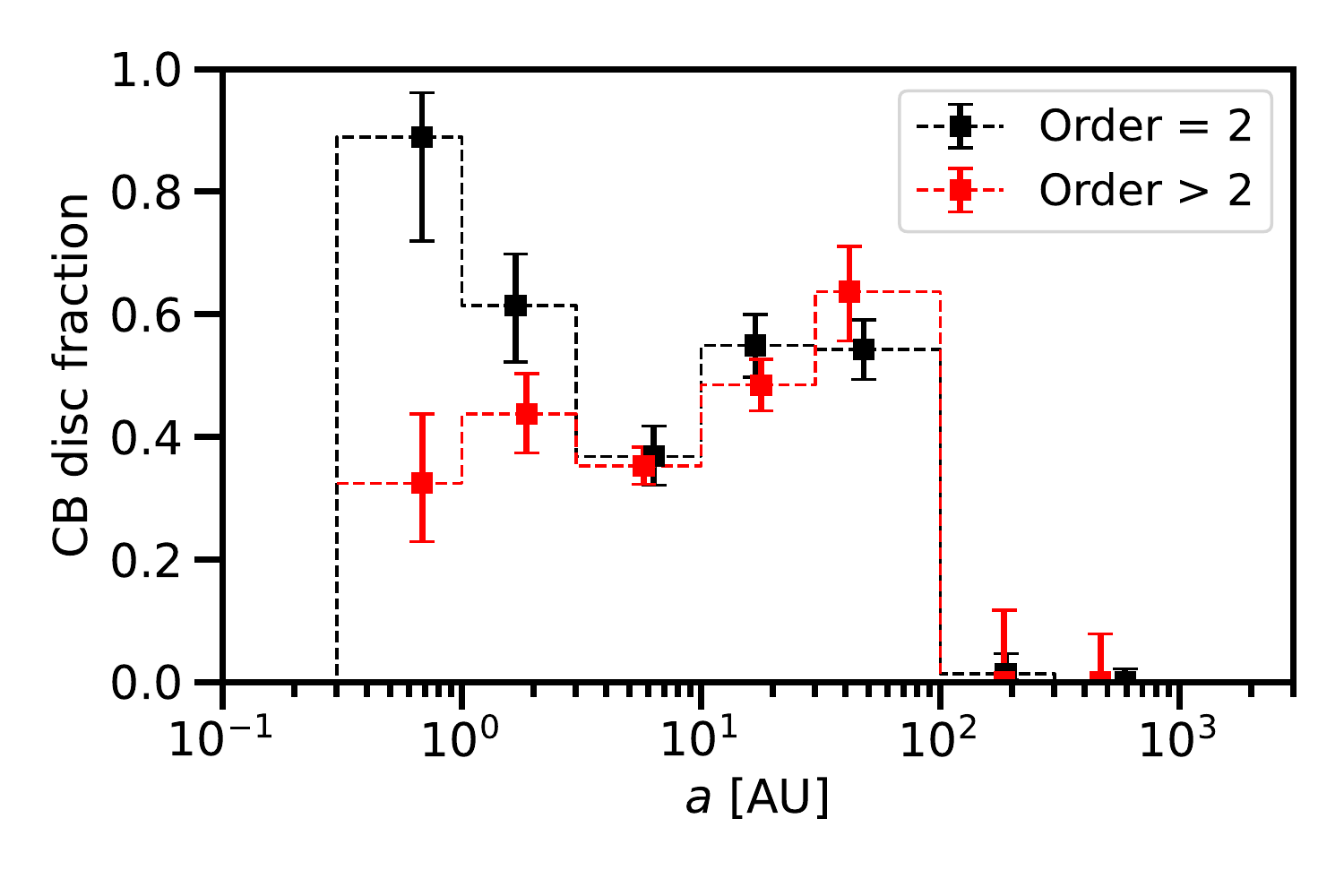} \vspace{-12pt}
    \caption{The fraction of binaries that host circumbinary discs plotted in bins of binary separation. The fraction of discs in pure binaries if plotted in black, and in red for hierarchical systems. The separation bins are divided as following: 0.3-1, 1-3, 3-10, 10-30, 30-100, and 300-1000; all in au. There are no discs in systems with separation larger than 300 au.}
    \label{fig:cbd_fraction}
\end{figure}

\begin{figure}
    \centering
    \includegraphics[width=0.98\linewidth]{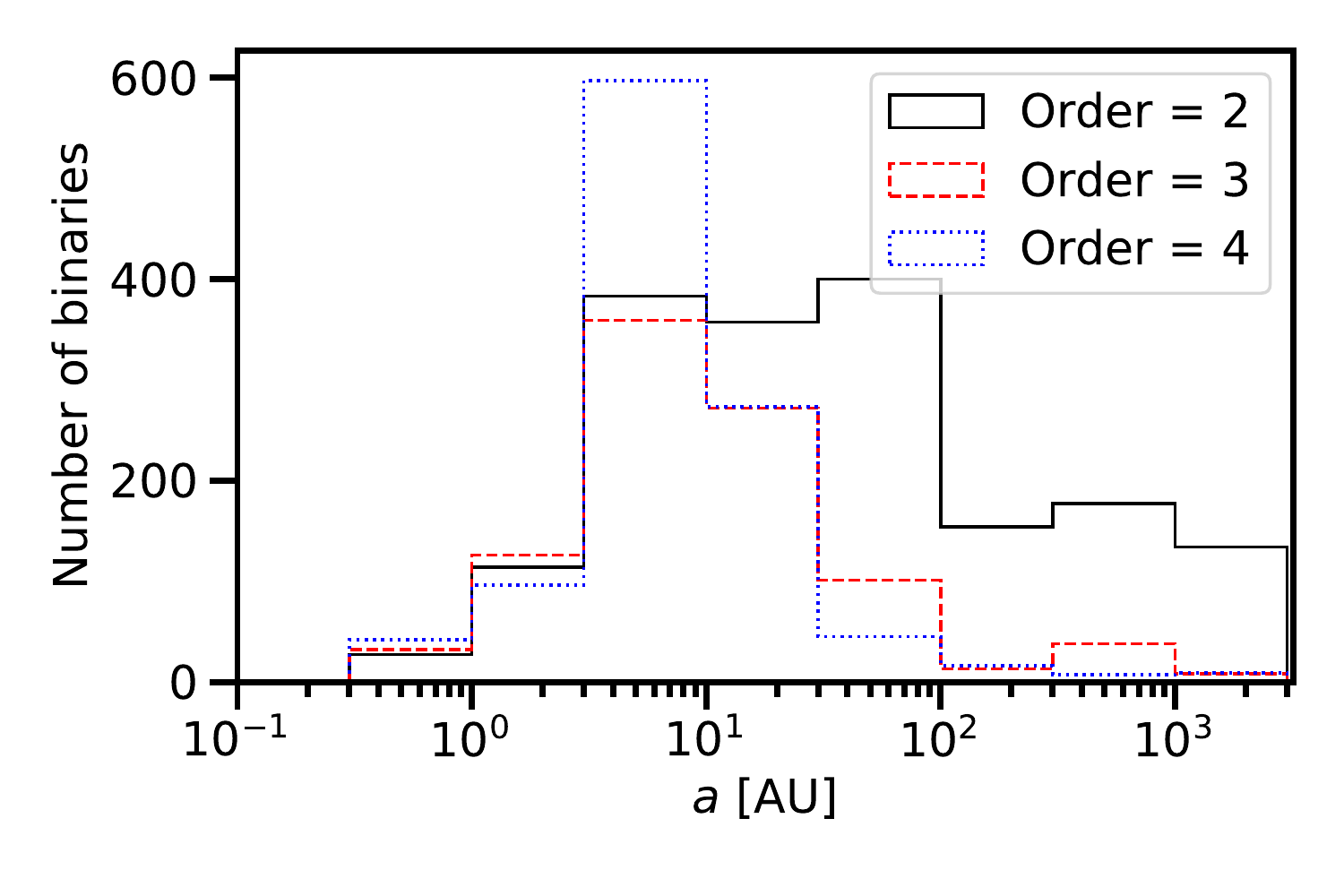} \vspace{-12pt}
    \caption{Instances of binary systems in bins of binary separation. We plot the number of binaries in each type of system, binary (solid black), and hierarchical triple (dashed red) and quadruple (dotted blue). The separations are binned as following: 0.3-1, 1-3, 3-10, 10-30, 30-100, 100-300, 300-1000, 1000-3000; all in au. There are no binaries with separations larger than 3000 au.}
    \label{fig:binstances}
\end{figure}

\subsection{Frequency of discs}
\label{sec:freq of discs}

Here we consider the fraction of binary systems that host a circumbinary disc. We consider pure binaries and binary pairs in hierarchical systems separately. The separations of the binaries are logarithmically binned every 0.5 dex, and then we count the number of discs about binaries in each bin. The resulting distributions are plotted in Fig. \ref{fig:cbd_fraction}. The points plotted in Fig.  \ref{fig:cbd_fraction} are the mean separation of the disc hosting binaries in each bin, and the associated error-bars are calculated using Wilson's interval \citep[][]{wilson_1927}, where the observed probability of a binary having a disc is given by $\hat{p} = (N_\mathrm{discs}+1)/(N_\mathrm{binaries}+2)$. The fractions are based upon the instances of binary pairs. We count the number of instances of pairs and the number of pairs with a CB disc across all snapshots and bin them by binary separation. When considering the fraction of pure binaries that host a CB disc we first note a bimodal distribution of CB disc fraction with peaks at $a \approx 0.7$ au, and $a \approx 30$ au in binary semi-major axis. The peak at smaller separations is higher than the peak at larger separations, with $89$ per cent of tight binaries hosting a disc compared to around $55$ per cent of intermediate separation binaries. 
Broadly the fractions of binaries in hierarchical systems that host a CB disc are similar except for tight binaries ($a \lesssim 3$ au); for $a > 3$ au there is no significant difference. In hierarchical systems tight binaries are less likely to host a CB disc than strict binary pairs. Note, $a$ in hierarchical systems is the separation between the closest bound pair in the system.

In Fig. \ref{fig:binstances} we show the number of binary system instances in the same separation bins that were used to calculated the CB disc fractions. For separations of $a<3$ au the number of binary systems are similarly distributed regardless of order. In hierarchical systems the number of binary pairs peaks in the $3<a<10$ au separation bin, and then tails off as separation increases with very few pairs with $a>100$ au. In pure binary systems the number of binaries is roughly uniform in log-separation for $3<a<100$ au, then the number of binaries decreases when $a>100$ au.

The range $3-100$ au has the highest total number of binaries (Fig. \ref{fig:binstances}).  Circumbinary disc hosting binaries with separations $3<a<100$ au tend to become more common as the separation increases up to $a\approx 100$ au (Fig. \ref{fig:cbd_fraction}); the CB disc fraction drops to almost zero when $a>100$ au in all systems. This suggests that binaries this wide are very rarely able to form a rotationally supported disc. When taking into account that the binary separation distribution peaks at $\approx 30$ au, both in the calculation and observations \citep[e.g.,][]{raghavan_multiplicity_2010ApJS..190....1R}, CB discs should typically be found around $10-100$ au binaries. 

For pure binaries, systems with $10<a<30$ au, and $30<a<100$ au roughly host the same fraction of discs. The error-bars give a $2\sigma$ confidence interval on the probability of a binary hosting a disc given the number of discs and binaries. We note the caveat that the sample of discs is not strictly independent.

\subsection{Disc statistics}
\label{sec: acc disc stats}

\begin{figure}
    \centering
    \adjincludegraphics[width=0.98\linewidth]{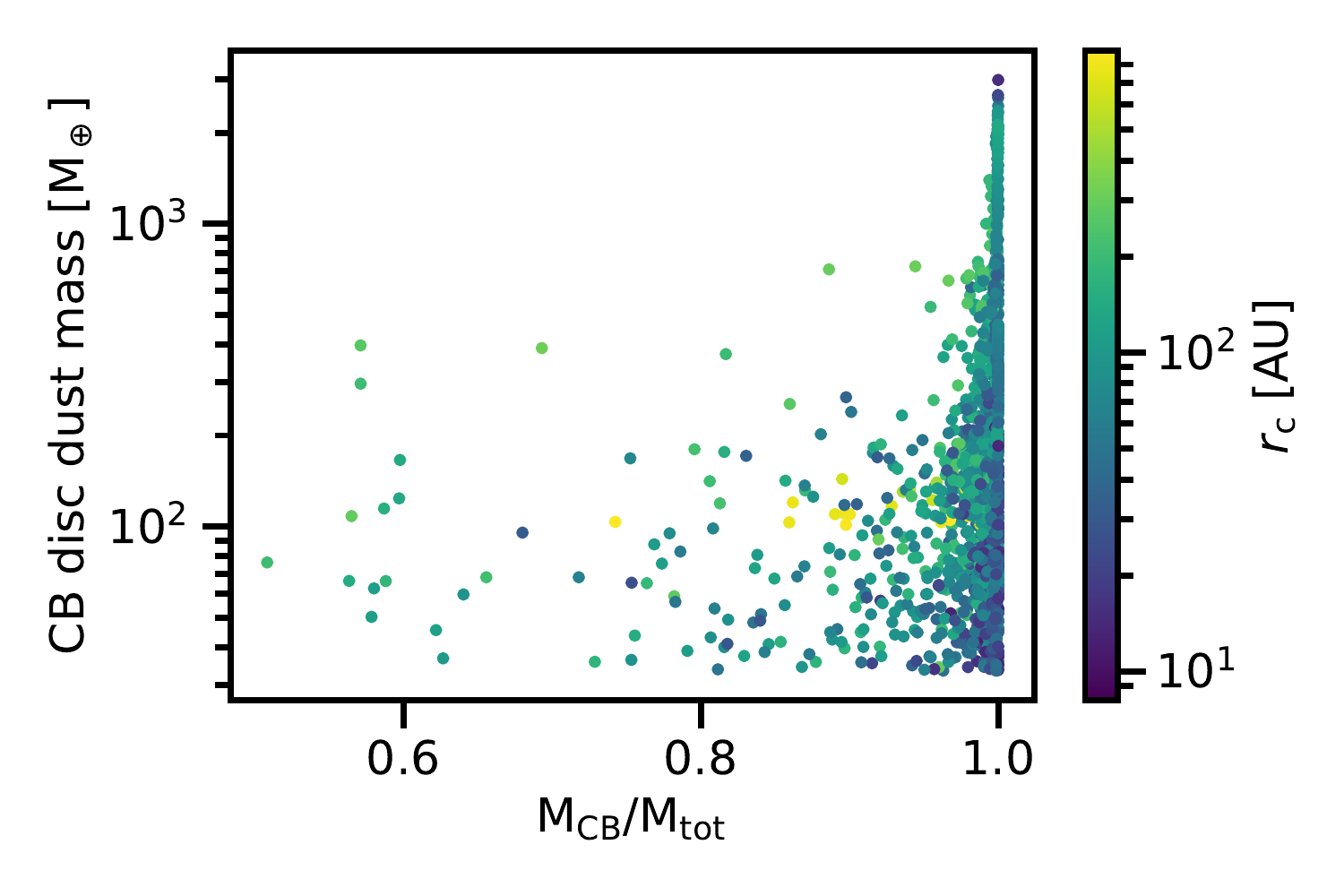} \vspace{-0.3cm}
    \caption{Graph of the circumbinary disc dust mass against the fraction of the circumbinary disc mass to the total disc mass.  In addition, the radius of the CB disc is shown using the colour of the points. The bulk of the CB discs contain more than 90 per cent of the total disc mass in of the binary. There are only a handful of systems for which more then 20 per cent of the disc mass is in the circumstellar discs. Typically, the lower the CB disc mass, the more likely it holds less of a proportion of the total disc material.}
    \label{fig:cb_disc_mass_fraction}
\end{figure}

In Fig. \ref{fig:cb_disc_mass_fraction} we plot the CB disc dust mass against the ratio of the CB disc mass to total system disc mass, with the characteristic radius of the CB discs shown by the colour of the points. Note that all plots in the remainder of the paper are generated from instances of CB discs and/or bound pairs. We assume a dust-to-gas ratio of 0.01 to compute the dust masses. As mentioned in Section \ref{sec:cb disc char} we only consider CB discs that contain at least $50$ per cent of the total disc mass. This may seem like a high percentage but considering $\sim 87$ per cent of CB discs contain more than $95$ per cent of the disc mass of the system it captures the vast majority of CB discs. The `discs' that are discarded with this criterion tend to be of a mass lower than $0.03~\mathrm{M}_\odot$ ($100~\mathrm{M}_\oplus$) and have a radius of less than the separation of the binary orbit. This suggests that young binary systems with CB discs tend to have significant discs, and that the discs about the component stars tend to be comparatively low mass. Discs with masses of less than $0.01 ~\mathrm{M}_\odot$ ($\approx 30 ~\mathrm{M}_\oplus$) are poorly resolved by the calculation, due to too few  ($<700$) SPH particles modelling them. 

\begin{figure}
    \centering
    \includegraphics[width=0.98\linewidth]{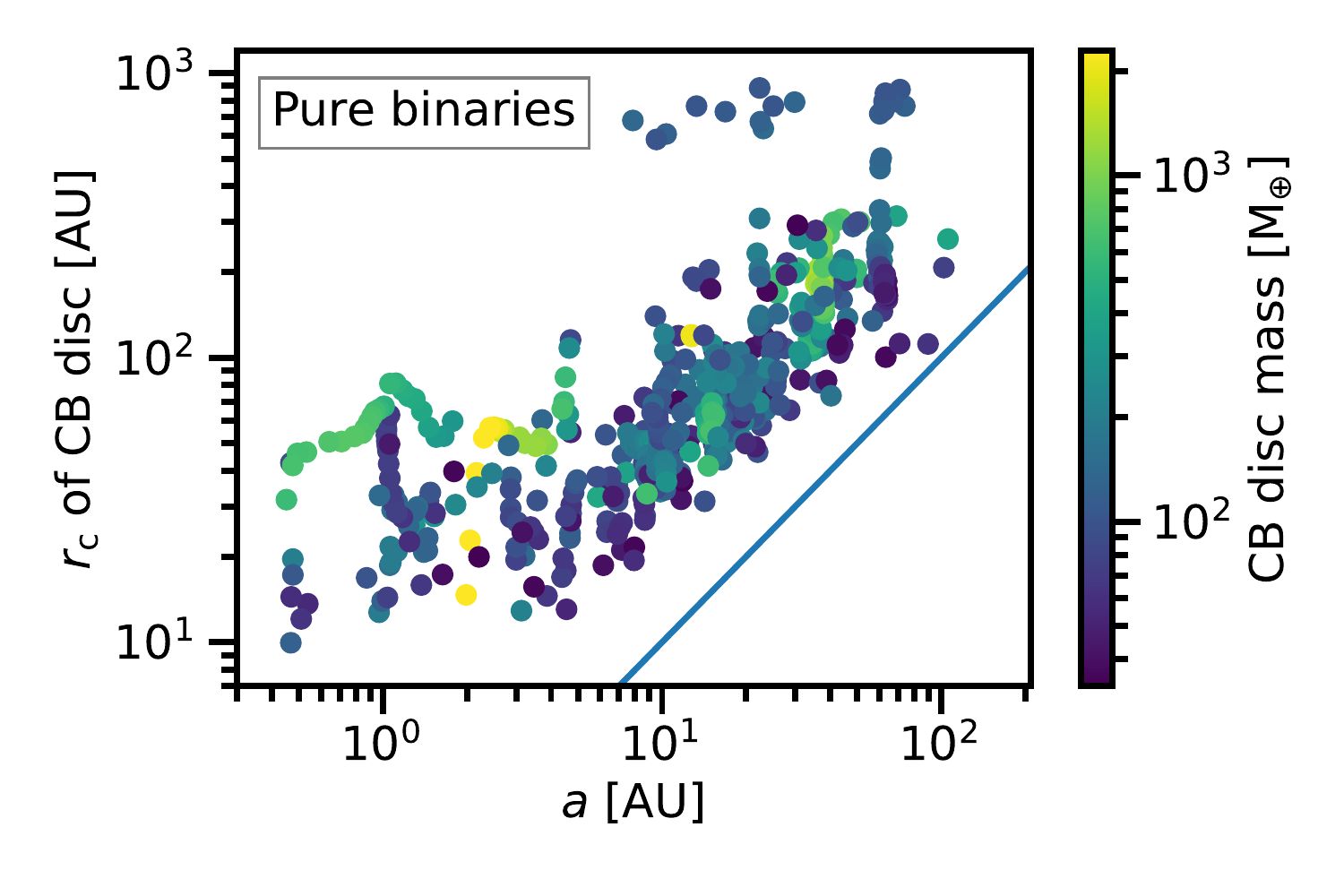}\vspace{-0.3cm}
    \includegraphics[width=0.98\linewidth]{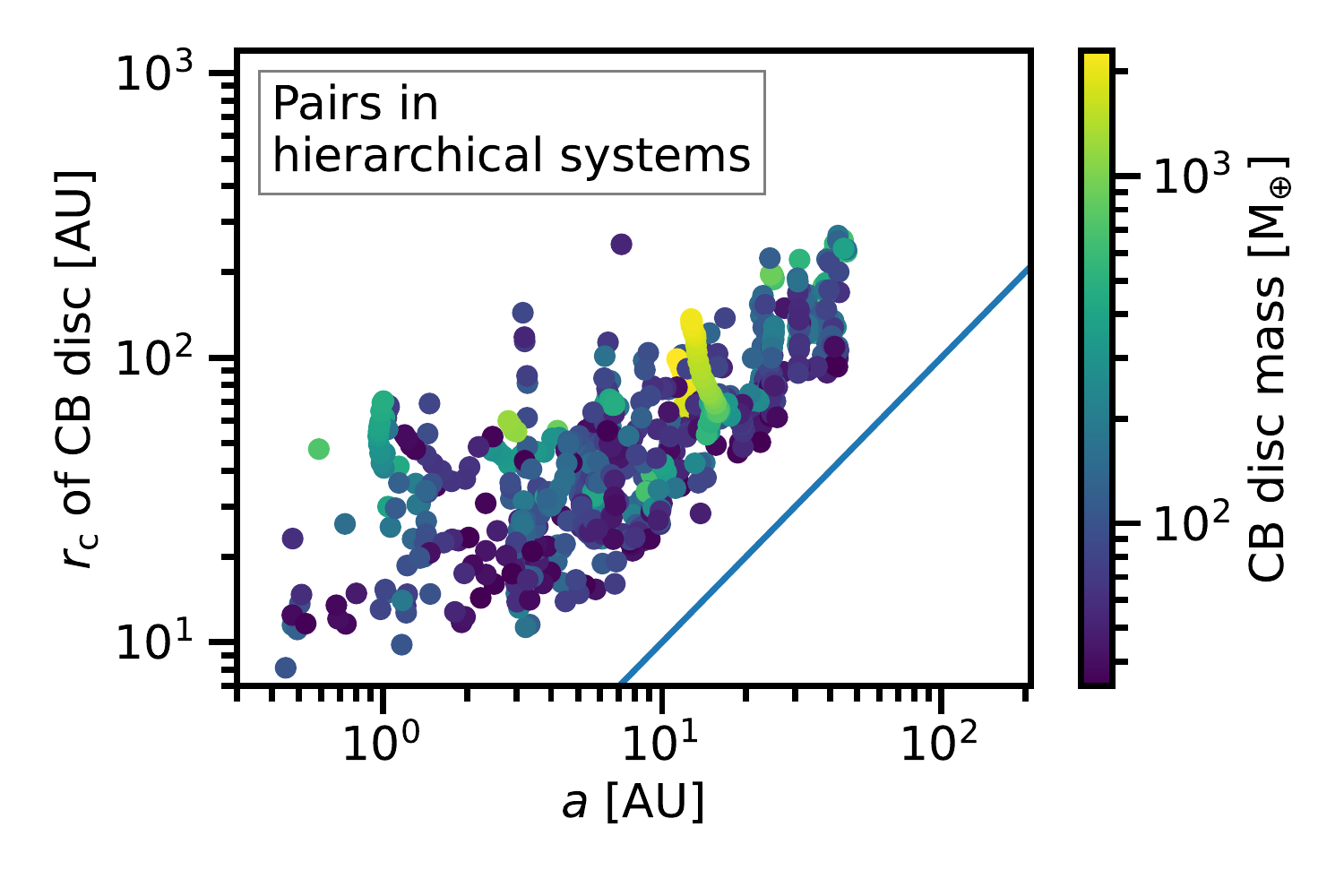}
    \caption{The radii of CB discs in pure binaries (top) and hierarchical systems (bottom), and how they depend on binary semi-major axis. The masses of the CB discs are shown using a colour map. The blue line in both plots is where r$_\mathrm{c} = a$. As expected, all CB disc radii are greater than the semi-major axis, typically by factors of 3-8 for $a\gtrsim 5$ au.}
    \label{fig:cb_disc_sep_rad_mass}
\end{figure}

\begin{figure}
    \centering
    \includegraphics[width=0.98\linewidth]{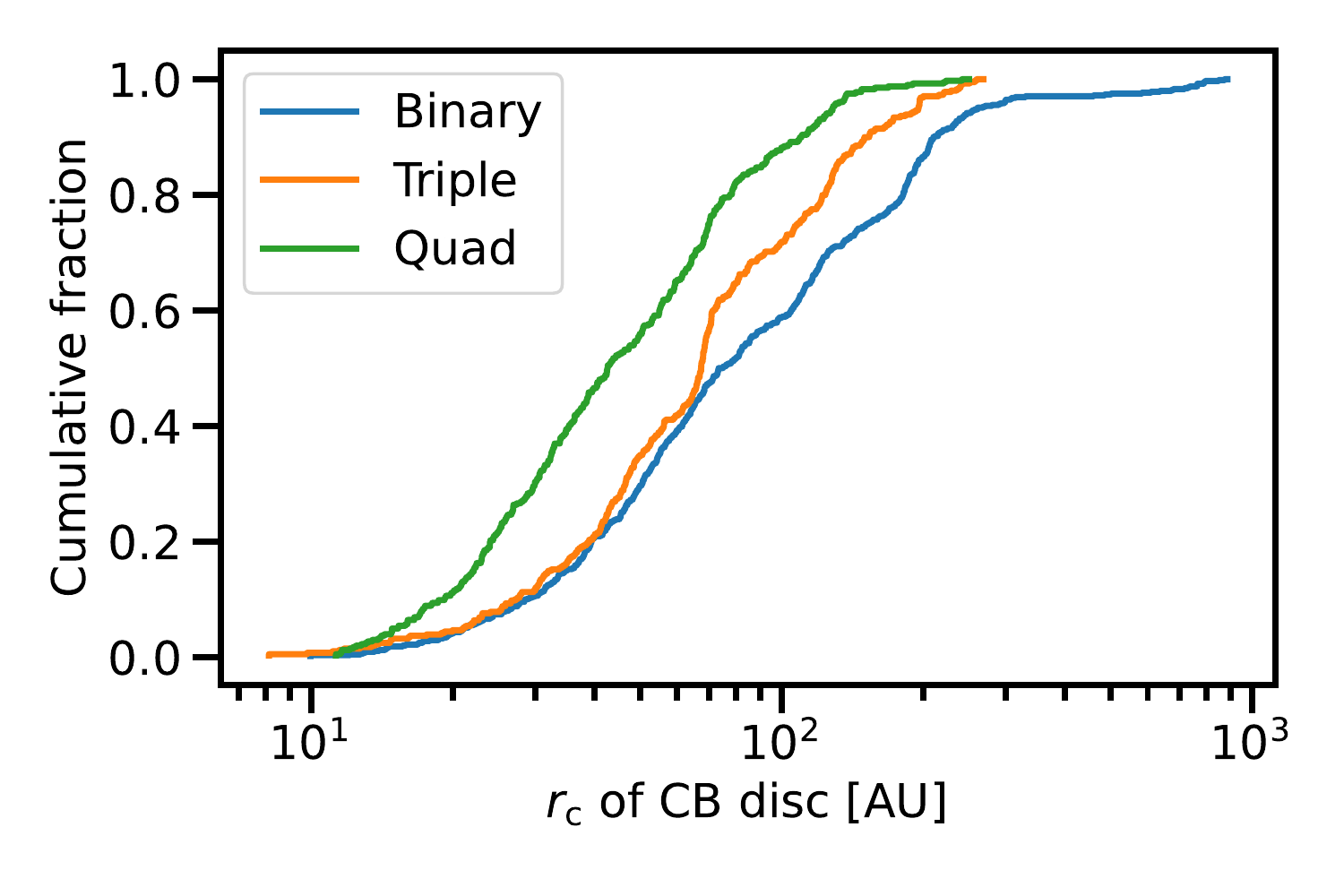}\vspace{-0.45cm}
    \includegraphics[width=0.98\linewidth]{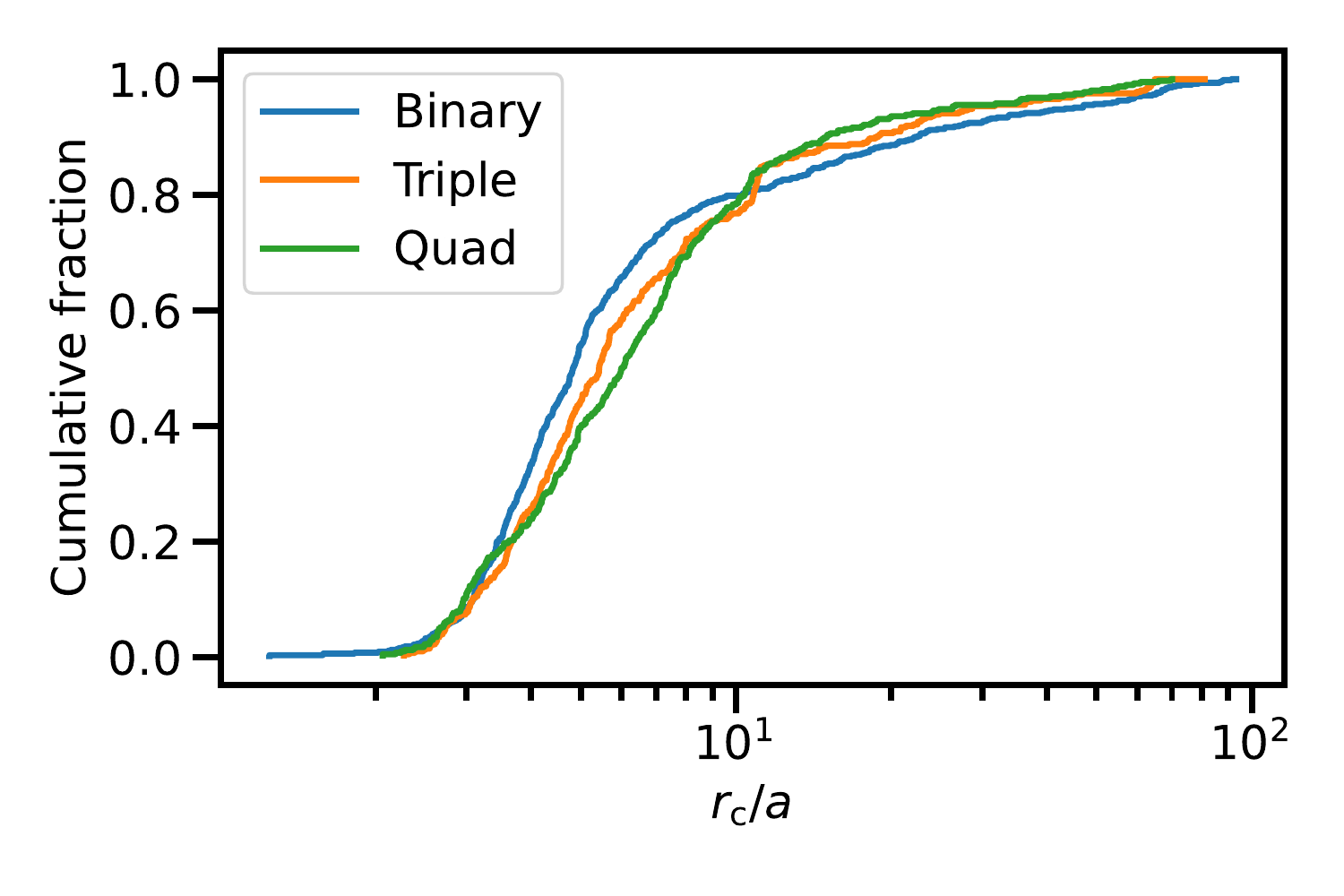}\vspace{-0.45cm}
    \includegraphics[width=0.98\linewidth]{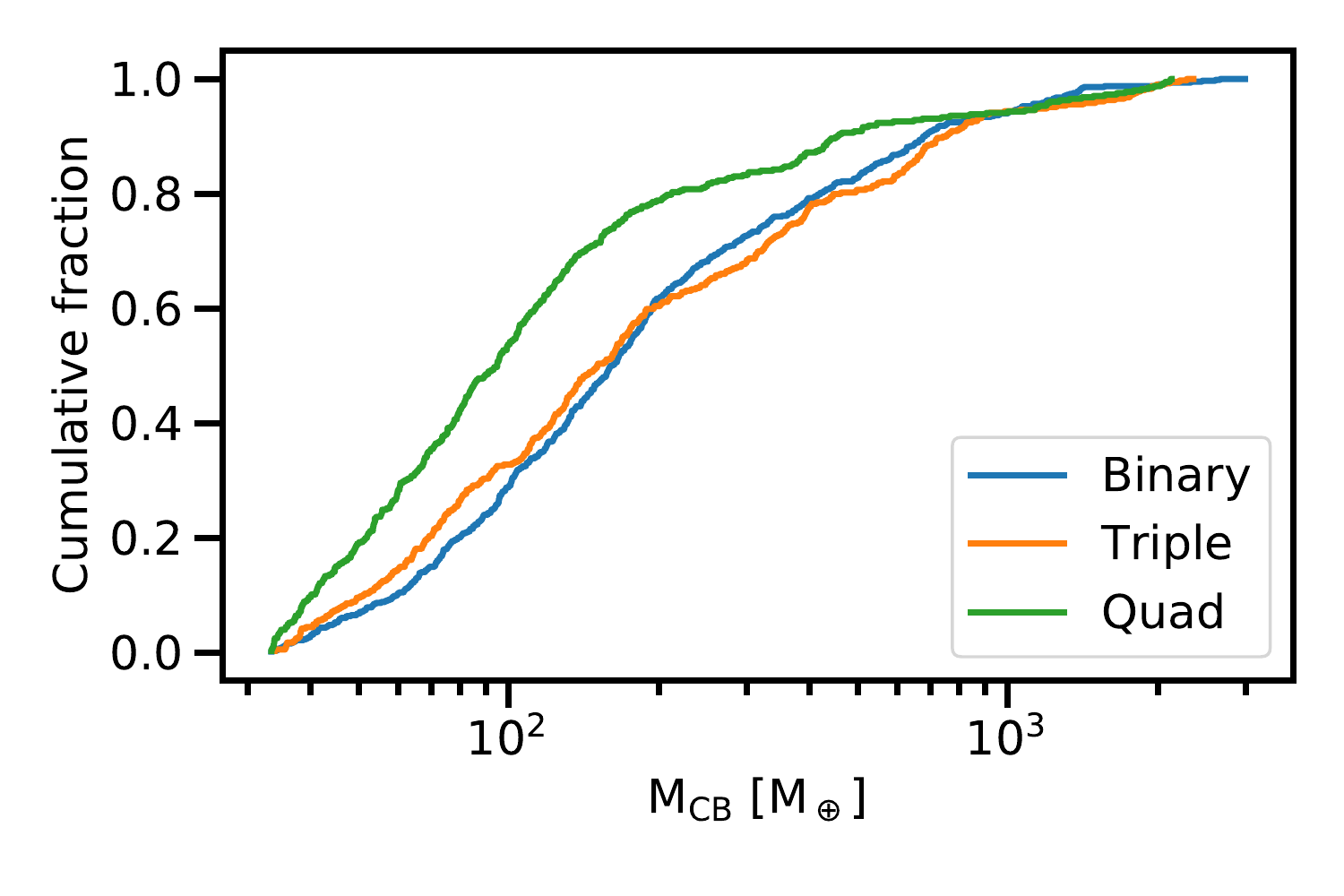}\vspace{-0.45cm}
    \includegraphics[width=0.98\linewidth]{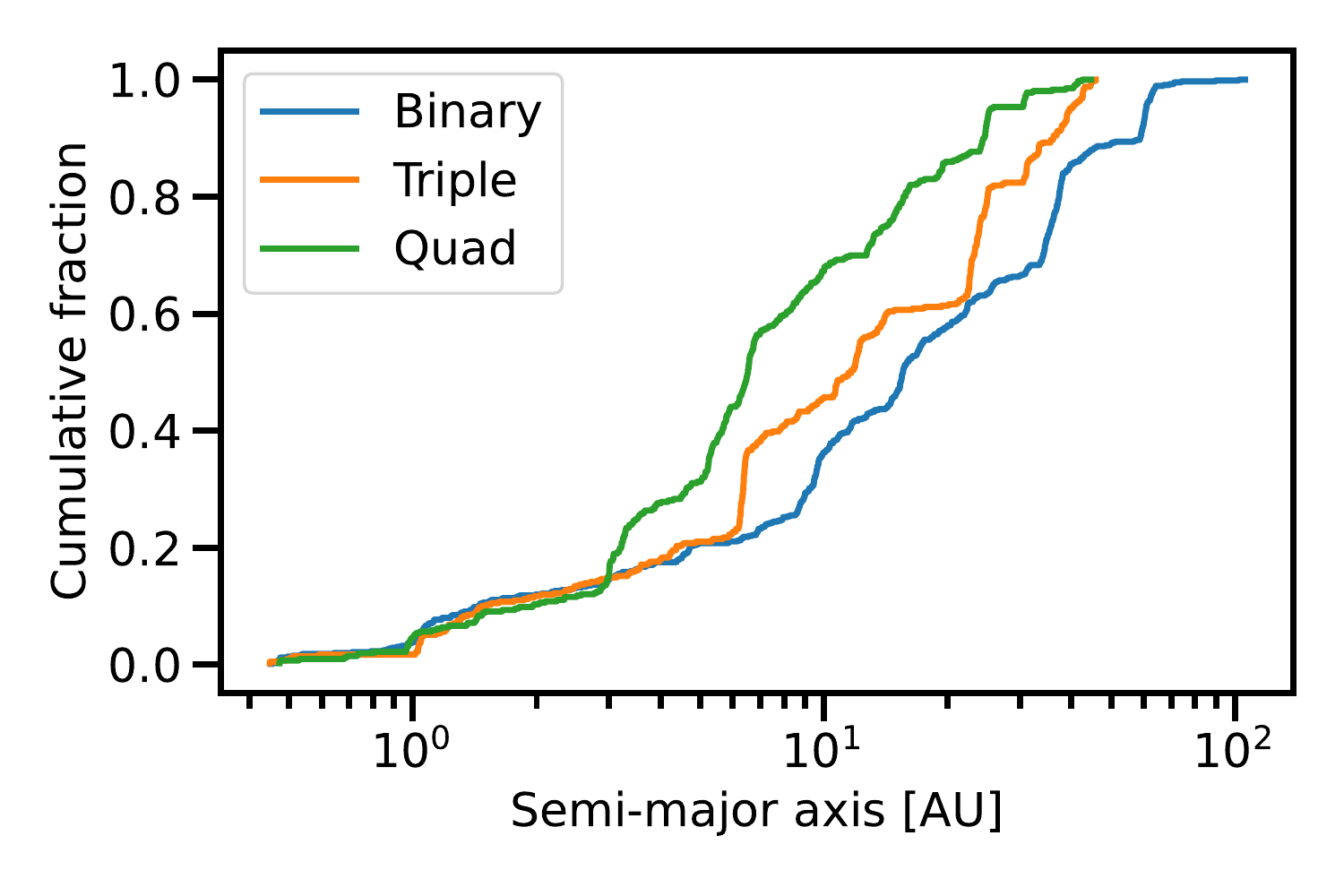}\vspace{-0.45cm}
    \caption{The cumulative distribution of CB disc radii measured in au (top) and as a fraction of the binary's semi-major axis (upper middle) in binary, triple, and quadruple systems.  We also give the cumulative distributions of CB disc dust masses (lower middle), and semi-major axes of bound pairs in pure binaries and hierarchical systems with CB discs (bottom). The radii of CB discs is lower systematically with increasing order, mainly because the binaries in higher-order systems are systematically closer.  The masses of CB discs in pure binaries and in hierarchical triples are very similar, but they are lower in hierarchical quadruples. Semi-major axis of bound pairs tends to decrease with increasing order.}
    \label{fig:cdf_mass_rad_multi}
\end{figure}

When considering the separation of the binary system we find that the radius ($r_\mathrm{c}$) of the CB disc tends to increase with separation.  We show this in Fig. \ref{fig:cb_disc_sep_rad_mass} (top panel) for pure binaries,  with the mass of the CB disc shown in the colour map. All binaries that host a CB disc have a separation $a < 110$ au with discs generally smaller than $r_\mathrm{c} \lesssim 200$ au. We find the median value of CB disc radius in units of binary semi-major axis, $a$, is $\sim4.9$ in pure binaries, and $\sim5.4$ and $\sim6.1$ in triple and quadruple systems, respectively (i.e. there is no great difference). We note that we find no trend between CB disc mass and CB disc radius. 

In Fig. \ref{fig:cb_disc_sep_rad_mass} (bottom panel) we find a similar trend for binaries in higher-order systems, however the maximum radii of the CB discs appear to be lower. None of the CB discs have a radius $r_\mathrm{c} > 300$ au, and there are also no disc-hosting binaries in hierarchical systems with a separation $a>50$ au. The radii of CB discs in quadruple systems are systematically lower than those in pure binary systems, see Fig. \ref{fig:cdf_mass_rad_multi} (top panel). However, this is slightly misleading because whilst the CB discs in hierarchical systems tend to be smaller, the separations of the binaries also tend to be smaller. Therefore, in the second panel of Fig. \ref{fig:cdf_mass_rad_multi} we show the cumulative distribution of disc radii measured as a fraction of binary separation (i.e., $r_\mathrm{c}/a$). CB discs sizes relative to their binary in all systems are similarly distributed, and thus the disc sizes don't differ significantly for a given binary separation. 

We plot the cumulative distribution for CB disc masses in systems of differing order in the third panel of Fig. \ref{fig:cdf_mass_rad_multi}; CB discs in triple and binary systems have a very similar distribution of masses, whereas the CB discs in quadruple systems tend to be about a factor of two lower in mass. In the bottom panel of Fig. \ref{fig:cdf_mass_rad_multi} we show the cumulative distributions of binary semi-major axis for binaries in binary, triple, and quadruple systems. Until binaries become more separated than $\sim 3$ au the semi-major axes of binary pairs are almost identical across multiplicity. After this point we see a deviation in the distribution of separations. Around $35$ per cent of pure binaries instances found are close ($a<10$ au), with this percentage increasing with multiplicity; more than $60$ per cent disc bearing binary instances in quadruple systems are close binaries. The median separation of CB disc hosting binaries is $a \approx 11$ au; this falls below the separation of binaries that one might reasonably expect to be able to observe in optically thick young protostellar systems ($a \approx 50$ au) \citep{tobin_vla_2016}. 

\begin{figure}
    \centering
    \includegraphics[width=0.98\linewidth]{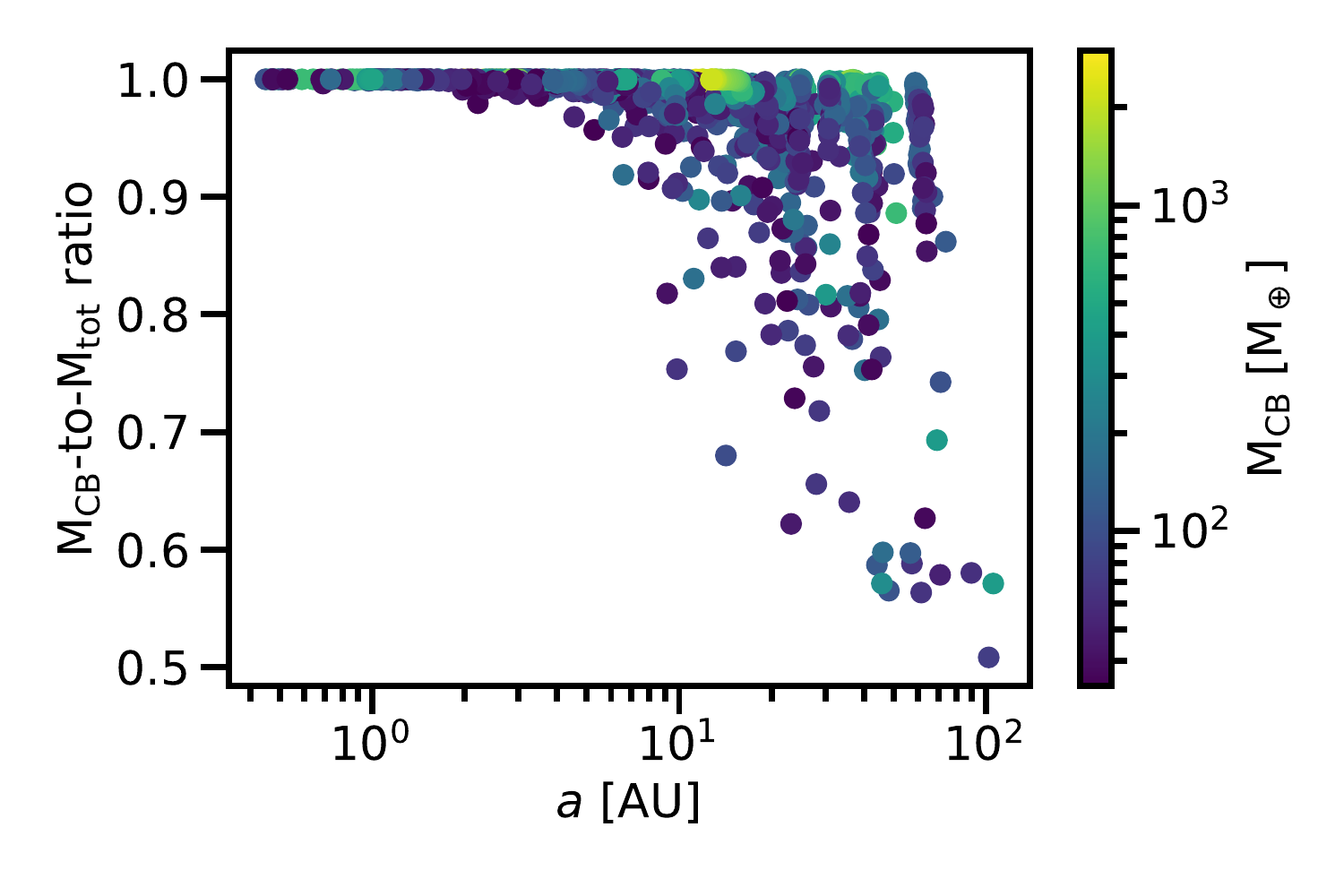} \vspace{-0.3cm}
    \caption{The circumbinary disc mass to system disc mass ratio plotted with the binary semi-major axis separation. CB disc mass in shown in the colour map. In systems with $a<10$ au most CB discs contain $>90$ per cent of total disc mass. As $a$ increases the proportion of total disc mass in the CB disc decreases.}
    \label{fig:cb_sep_massratio_mass}
\end{figure}

Whilst we find no correlation between binary separation and CB disc mass, we do find a trend between the ratio of CB disc to total disc mass and binary separation. We plot this in Fig. \ref{fig:cb_sep_massratio_mass}, with the CB disc mass shown in the colour map. CB disc mass dominates the total disc mass in the system for binaries with a separation $a\lesssim 10$ au.  As the binary separation increases past $a=10$ au, the CB disc to total system (CB including primary and secondary discs) disc mass ratio begins to have values below unity. This is in good agreement with observed binary systems \citep[e.g.][]{harris_taurus_multiple_systems_2012ApJ...751..115H}. However, in the simulation this is probably largely numerical. The sink particles modelling the protostars have accretion radii of 0.5~au, meaning that circumstellar discs smaller than a few au in radius are poorly modelled.  For a binary with a semi-major axis $a=10$~au, any circumstellar discs would be truncated to a few au in radius or even smaller for eccentric binaries.  Such small circumstellar discs would not be numerically resolved.

\begin{figure}
    \centering
    \includegraphics[width=0.98\linewidth]{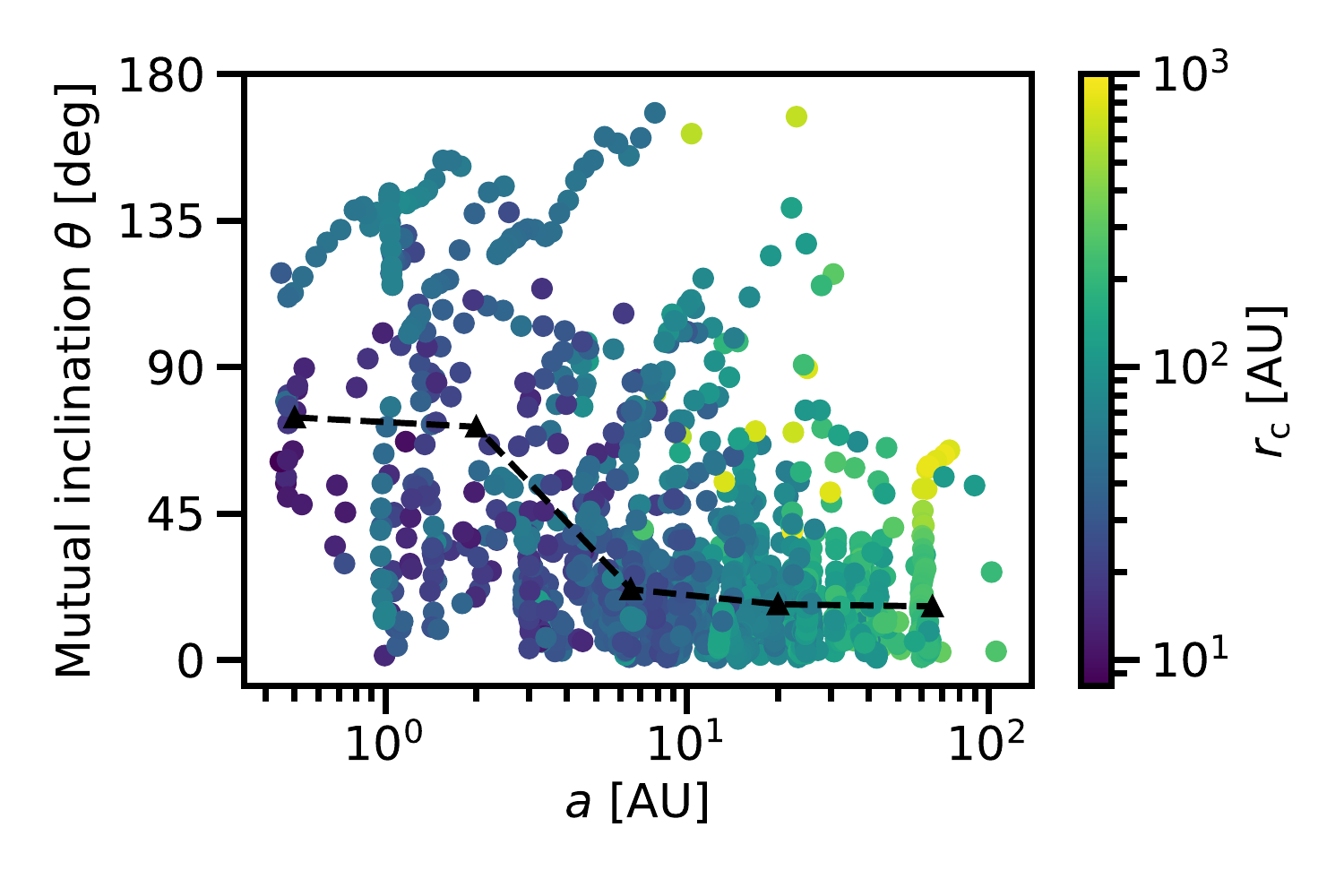} \vspace{-0.3cm}
    \caption{The mutual inclination angles of circumbinary discs relative to their binary's orbital plane versus binary orbit semi-major axis. The circumbinary disc radius is plotted in the colour map. The median mutual inclinations of discs binned every 0.5 dex are shown as black triangles. The discs about the very close binaries have a greater range of mutual inclinations.  These discs also tend to be relatively small, on the order of 10's of au. These small discs may cast shadows on any outer disc material. }
    \label{fig:cb_sep_orient_ecc}
\end{figure}

In Fig. \ref{fig:cb_sep_orient_ecc} we plot the mutual inclination angle between the CB disc plane and binary orbit plane versus the semi-major axis of the binary, with the radius of the CB disc shown in the colour map. We see that wider binaries typically have a CB disc that is more well aligned with the binary's orbital plane. The close binaries ($a<10$ au) have a greater range of mutual inclinations relative to their CB disc. Additionally, these close binaries tend to be more eccentric than the wider binaries. Misaligned discs can cause disc breaking, as has been previously shown in 3D hydrodynamical simulations \citep[e.g.][]{nixon_tearing_2013,facchini_wave-like_2013}, and may have been observed in the GW Orionis system \citep[][]{kraus_triple-star_2020}. Highly mutually-inclined small CB discs, such as we find here, can be the cause of asymmetric shadows cast on to outer disc material as detected in scattered light observations \citep[e.g.][]{benisty_shadows_2017,casassus_inner_2018,price_circumbinary_not_transitional_2018MNRAS.477.1270P}.

\begin{figure}
    \centering
    \includegraphics[width=0.98\linewidth]{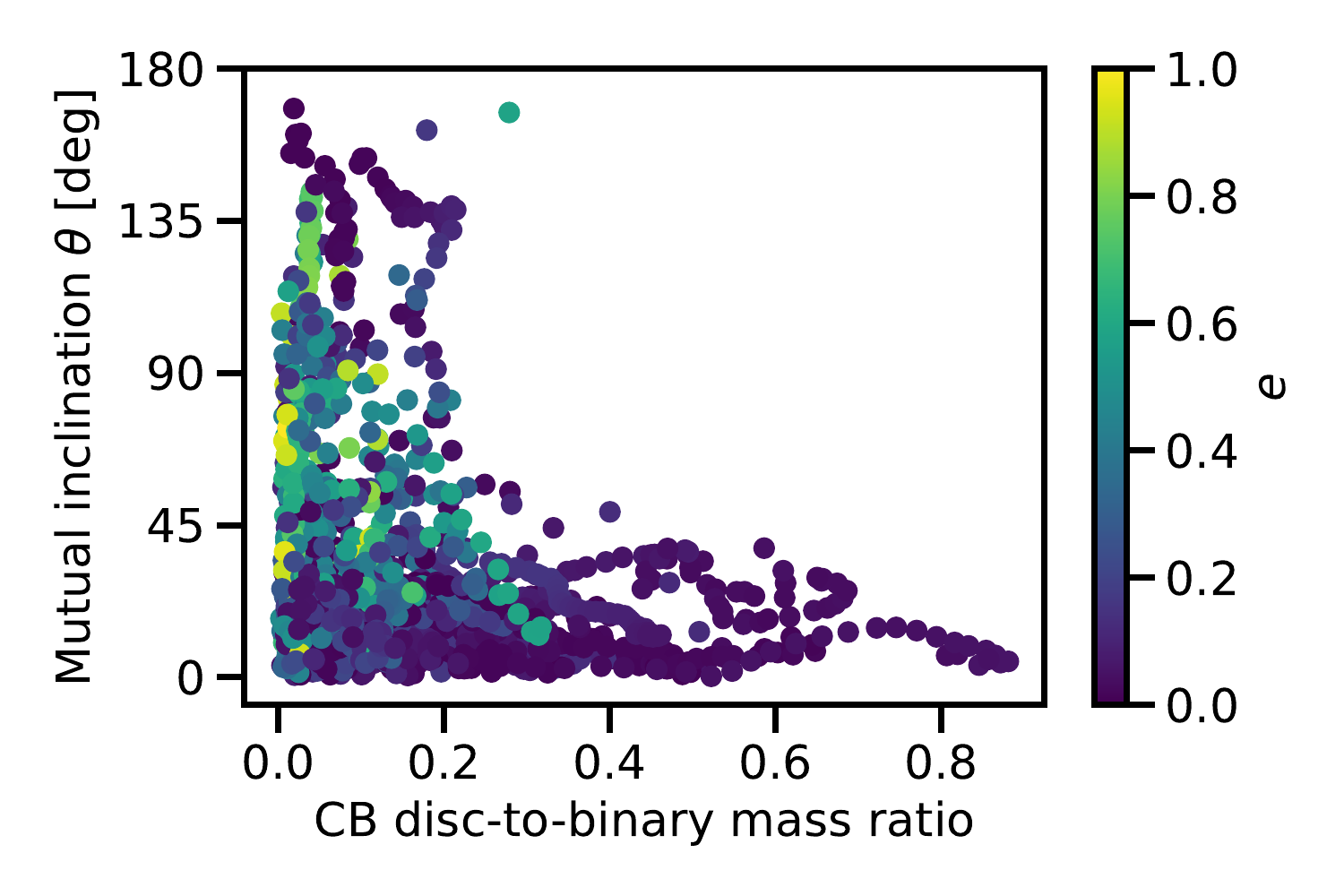} \vspace{-0.3cm}
    \caption{The mutual inclination angles of CB discs relative to their binary orbits versus the ratio of the CB disc gas mass to the mass of the binary.  The orbital eccentricity of the binary is plotted using the colourmap. }
    \label{fig:dsratio_orient_ecc}
\end{figure}

\citet{martin_lubow_polar_alignment_i_2017ApJ...835L..28M} found a polar alignment mechanism for CB discs that operates best at high binary eccentricity, and low disc mass. Also see \citet{lubow_linear_2018, zanazzi_inclination_2018}, and \citet{martin_polar_2019}. CB discs that are initially mildly misaligned with the binary orbit can evolve to a polar alignment, undergoing nodal libration oscillations of both tilt angle and the longitude of ascending node. They find this process operates above a critical angle of initial misalignment which depends upon the eccentricity of the binary and the mass of the disc. For binaries with eccentricity of 0.5, the process operates for discs with a mass that is just a few per cent of the binary mass with an initial misalignment of at least $40\degr$. In Fig. \ref{fig:dsratio_orient_ecc} we show disc to binary mass ratios and eccentricities of orbits against mutual inclination. We also find that the most misaligned discs have low disc masses relative to the binary's mass, and tend to have eccentric binaries, in agreement with \citet{martin_lubow_polar_alignment_i_2017ApJ...835L..28M}. CB discs whose masses exceed $30$ per cent of the binary mass typically have misalignment angles less than $40\degr$.

\section{Discussion}
\label{sec:discussion}

Here we focus the discussion on the mutual inclinations of the CB discs and their binary orbits, comparing those formed in the calculation with those reported in the literature. 

\subsection{Mutual inclinations}
\label{sec:comp with obs}

\begin{figure*}
\centering
\includegraphics[scale=2.25]{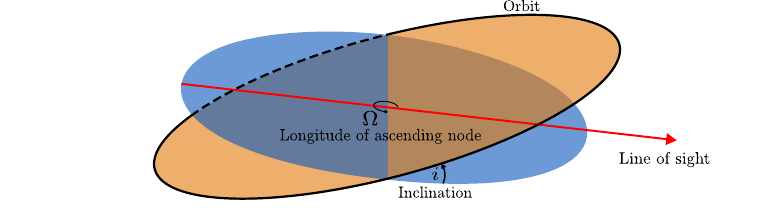}
\caption{A schematic showing angles of ascending node ($\Omega$) and inclination ($i$) of a body orbiting a central point with respect to the line of sight. The blue disc shows what the orbit would look like if it was in the plane of the observer's view, the orange disc is the orbit with an inclination to the observer's plane, and rotated around the centre point. Determining the angles of the ascending node and the inclination of both the binary and the disc are required to calculate the mutual inclination between the two. Due to the geometry of such systems, often there is a $180\degr$ ambiguity of the value of $\Omega$, i.e. it is hard to tell which way the system is 'facing'.}
\label{fig:orbit_geom}
\end{figure*}

The mutual inclination between a CB disc and the orbit of its binary, $\theta$, is defined as
\begin{equation}
\label{eq:mutual inclination}
    \cos{\theta} = \cos{i_\mathrm{disc}}\cos{i_*} + \sin{i_\mathrm{disc}}\sin{i_*}\cos{\left(\Omega_\mathrm{disc}-\Omega_*\right)},
\end{equation}
where $i_\mathrm{disc}$ is the inclination of the disc relative to the line of sight, $i_*$ is similar for the binary orbit, $\Omega_\mathrm{disc}$ is the longitude of the ascending node, and $\Omega_*$ is similar for the binary orbit. See Fig. \ref{fig:orbit_geom} for a sketch of the geometry. Calculating the mutual inclination between an observed CB disc and the orbit of its binary can be difficult. You need the orbital parameters of the binary, then you need good enough observations of the CB disc to find the inclination and the longitude of ascending node.

There are some known mutual inclinations between discs and binaries in the literature. HD 98800B is a near equal-mass binary ($q=0.86$) with a CB disc in a polar orbit with $\theta = 88.4 \degr \pm 2$ according to \citet{kennedy_circumbinary_2019} and $\theta = 92\degr \pm 3\degr$ as reported by \citet{czekala_degree_2019}. The disc around HD 142527 B has a mutual inclination that was reported to be $\theta = 35\degr \pm 5\degr$ \citep[][]{biller_likely_2012,lacour_m-dwarf_2016,boehler_close-up_2017,price_circumbinary_not_transitional_2018MNRAS.477.1270P, claudi_sphere_2019,czekala_degree_2019}. However, with improved orbital parameters \citet{balmer_improved_2022} suggest $\theta \approx 46 \degr \pm 2\degr$ or $\theta \approx 76 \degr \pm 3\degr$ depending on the value used for the longitude of ascending node as there is a $180\degr$ ambiguity with this. Using precise measurements of RV time series and disc dynamical mass \citet{czekala_degree_2019} were able to infer the mutual inclination angles using a hierarchical Bayesian model for four CB discs; V4046 Sgr, AK Sco, DQ Tau, and UZ Tau E, with inferred $\theta$ $<2.3\degr$, $<2.7\degr$, $<2.7\degr$, and $<2.7\degr$, respectively. Using the orbital parameters given by \citet{long_architecture_2021}, V892 Tau has $\theta = 5\degr \pm4\degr $. Another notable mention is the disc around V773 Tau B, which is thought to be in a polar alignment with its binary given the geometry of the system as a whole, but no value for $\theta$ has been reported \citep[][]{kenworthy_eclipse_2022}.

\begin{figure}
    \centering
    \includegraphics[width=0.98\linewidth]{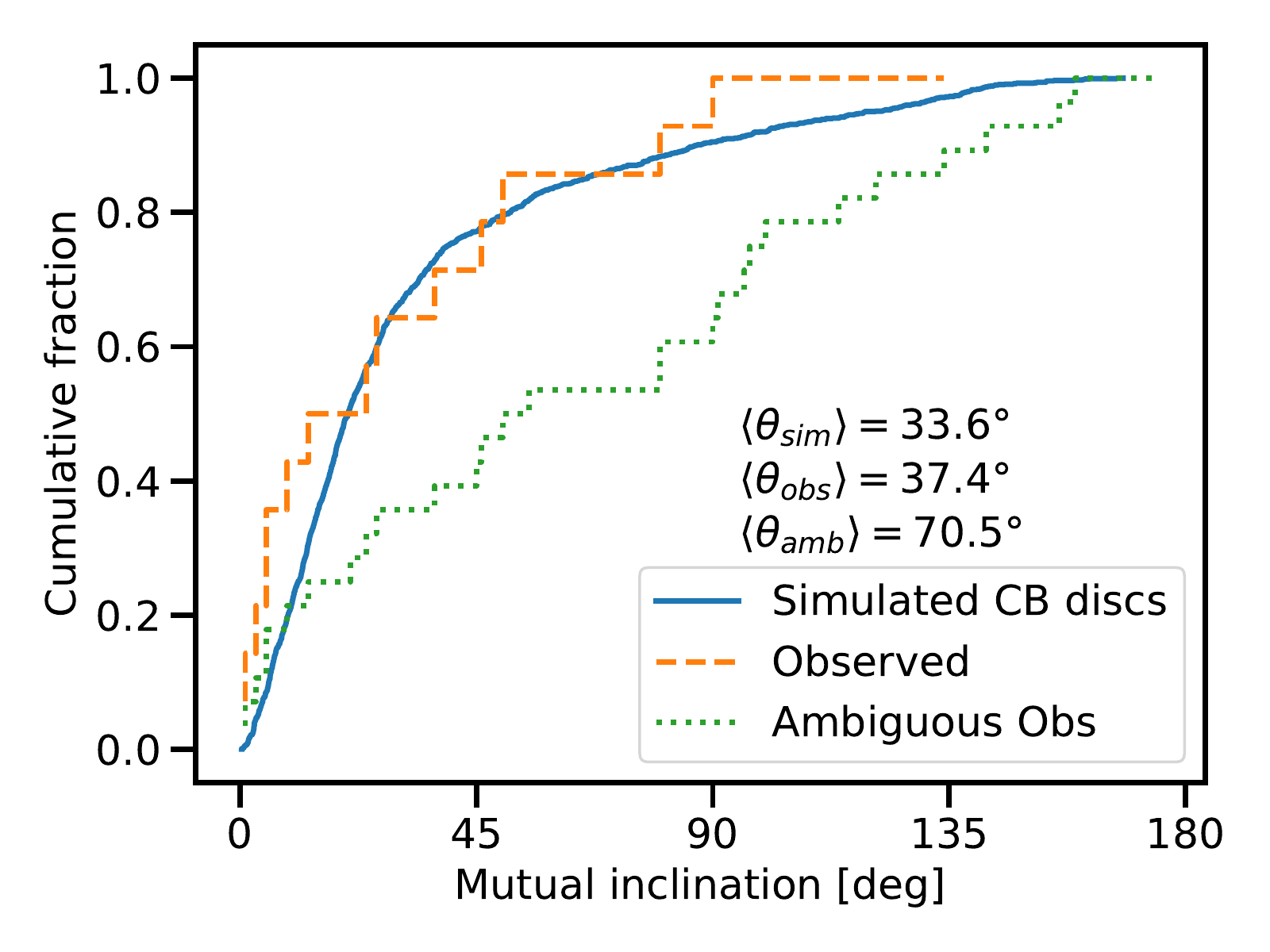}
    \caption{Cumulative distributions of the mutual inclination of circumbinary discs with their binaries as formed in the solar metallicity calculation of \citet{bate_metallicity_2019MNRAS.484.2341B}, and the mutual inclinations of observed discs from Table \ref{tab:mutual_incs}. The reported mutual inclinations ($\theta_1$, orange dashed), and both the reported `fake' values ($\theta_2$) we calculate due to the ambiguity in $\Omega_*$. The mean values of the mutual inclinations for the simulated, reported and ambiguous are $33.6\degr$, $37.4\degr$, and $70.5\degr$, respectively.}
    \label{fig:theta}
\end{figure}

Often when observing CB discs there is a $180\degr$ ambiguity with the longitude of ascending node, $\Omega$, for both the disc, $\Omega_{\text{disc}}$, and the binary $\Omega_*$. From Fig. \ref{fig:orbit_geom}, it is apparent that the geometry of the system will look the same from the observer's point of view if the disc is rotated $180\degr$ about the central node.  Additional information, such as radial velocities (RVs) or astrometry data, is required to eliminate this ambiguity. In the absence of this, there are two possible values of the mutual inclination between the binary orbit and the circumbinary disc. When studying the relative orientations of triple systems \citet{sterzik_relative_2002} note this ambiguity and rather than using one value (even if there are possibly two) to generate statistics they use both values; one "true" and one "false". They do this even for systems that have very well constrained orbital parameters. They find that the mean value of the mutual inclinations does not change, however we believe this is because the distribution of mutual inclination angles for triple systems is close to random. For a set of data that are not random the inclusion of the "false" values will change the mean value.  We give mutual inclination results from the calculation both using the actual values of $\theta$, as well as using two values of mutual inclination for each CB disc/binary orbit, since there are only a few observed discs that have a well constrained angle of ascending node.

We use the mutual inclinations of observed CB discs to produce a cumulative distribution to compare with the mutual inclinations of CB discs from our calculations. We provide the data for observed discs, along with the necessary literature values to calculate $\theta$, in Table \ref{tab:mutual_incs}. The literature values are given as $\theta_1$, and the values taking into account the ambiguity in $\Omega_*$ that we calculate are given as $\theta_2$. We note that we add $180\degr$ to $\Omega_*$ to obtain the ambiguous values; this choice is arbitrary. We chose to use $\Omega_*$ as the value with ambiguity as in the literature there are more systems with ambiguity attached to this value.  In Appendix \ref{AppendixA} we provide some extra discussion of how we chose the values for some of the observed systems in Table \ref{tab:mutual_incs}.

\subsubsection{Comparison of simulated and observed mutual inclinations}

We produce a cumulative distribution for the simulated discs, reported values for observed discs, and reported and ambiguous values combined for observed discs and plot them in Fig. \ref{fig:theta}. The mutual inclinations of the simulated CB discs have a very similar distribution to the reported observed mutual inclinations, with around half of all CB discs being inclined by less than $25\degr$ to the binary's orbit. At around $40\degr$ the distribution begins to flatten out, with $\approx 20$ per cent of CB discs having an inclination of $>45\degr$. Of the simulated CB discs, $\approx 10$ per cent are retrograde (inclined more than $90\degr$ to the binary orbit), compared to $5$ per cent of the reported observed discs. When combining the reported values with the alternate values the distribution of mutual inclinations approaches a more random, uniform distribution, with around $40$ per cent of discs possibly being retrograde.

Using a hierarchical Bayesian analysis, \citet{czekala_degree_2019} suggest that $\approx 70$ per cent of short period spectroscopic binaries have $\theta < 3\degr$, whereas objects with longer orbital periods ($> 30$ days) have a much greater range of inclinations, from coplanar to polar. In relation to the calculation discussed in this paper, very close spectroscopic binaries are not produced due to the accretion radii of the sink particles $r_\mathrm{acc} = 0.5$ au. We do not see such a degree of coplanarity of CB disc and binary orbit in the closest binaries produced by the simulation (see Fig. \ref{fig:cb_sep_orient_ecc}).  This may be due to the limited gas resolution of the calculation and/or the fact that the simulated systems are all very young (ages $<10^5$ years) and haven't yet had time to become aligned through the action of gravitational torques and accretion \citep[e.g.,][]{bate_chaotic_star_formation_2010MNRAS.401.1505B}.

\begin{table*}
    \newcounter{foo}
    \centering
    \begin{tabular}{ccccccccccccccc}
        \hline
        \hline
        Name  & $\theta_1$ & $\theta_2$ & $i_\mathrm{*}$ & $i_\mathrm{disc}$ &  $\Omega_\mathrm{*}$ &  $\Omega_\mathrm{disc}$ &  $\Omega_\mathrm{*, amb}$ &  References \\
         & [deg] & [deg] & [deg] & [deg] & [deg] & [deg] & [deg] & \\ \hline
        TWA 3A & 9 $\pm$ 6 & 97 $\pm$ 1 & 48.5 $\pm$ 0.8 & 48.8 $\pm$ 0.7 & 104 $\pm$ 9 & 116 $\pm$ 0.4 & 284 $\pm$ 9 & (\ref{ref1}), (\ref{ref2}), (\ref{ref3}) \\ 
        AK Sco & 3 $\pm$ 2 & 142 $\pm$ 2 & 108.8 $\pm$ 2.4 & 109.4 $\pm$ 0.5 & 48 $\pm$ 3 & 51.1 $\pm$ 0.3 & 228 $\pm$ 3 & (\ref{ref1}), (\ref{ak_sco_anthonoiz}), (\ref{ak_sco_czekala}) \\ 
        HD $131511^\mathrm{a}$ & 5 $\pm$ 9 & 175 $\pm$ 9 & 93.4 $\pm$ 4.2 & 90 $\pm$ 10 & 248 $\pm$ 3.6 & 245 $\pm$ 5 & 68 $\pm$ 3.6 & (\ref{hd_kennedy_2015})  \\ 
        alpha CrB$^\mathrm{a}$ & 24 $\pm$ 28 & 156 $\pm$ 28 & 88.2 $\pm$ 0.1 & 90 $\pm$ 10 & 330 $\pm$ 20 & 354 $\pm$ 20 & 150 $\pm$ 20 & (\ref{alpha_kennedy_2012}) \\ 
        beta Tri$^\mathrm{a}$ & 1 $\pm$ 8 & 100 $\pm$ 10 & 130 $\pm$ 0.5 & 130 $\pm$ 10 & 245.2 $\pm$ 0.67 & 247 $\pm$ 10 & 74.5 $\pm$ 0.67 & (\ref{alpha_kennedy_2012})  \\ 
        HD 98800B & 134 $\pm$ 3 & 91 $\pm$ 3 & 67 $\pm$ 3 & 154 $\pm$ 1 & 343 $\pm$ 2.4 & 196 $\pm$ 1 & 163 $\pm$ 2.4 & (\ref{hd_98800b_kennedy_2019}), (\ref{hd_98800b_zuniga-fernandez_2021}) \\ 
        GW Ori A-B & 50 $\pm$ 6 & 45 $\pm$ 6 & 157 $\pm$ 7 & 137 $\pm$ 2 & 263 $\pm$ 13 & 1 $\pm$ 1 & 83 $\pm$ 13 & (\ref{gw_ori_ab_czekala_2017})  \\ 
        HD 200775 & 13 $\pm$ 8 & 121 $\pm$ 7 & 66 $\pm$ 7 & 55 $\pm$ 1 & 172 $\pm$ 6 & 180 $\pm$ 8 & 352 $\pm$ 6 & (\ref{hd_200775_monnier}), (\ref{hd_200775_okamoto_2009}), (\ref{hd_200775_benisty}) \\ 
        V892 Tau & 5 $\pm$ 4 & 114 $\pm$ 3 & 59.3 $\pm$ 2.7 & 54.6 $\pm$ 1 & 50.5 $\pm$ 9 & 53 $\pm$ 0.7 & 230.4 $\pm$ 9 & (\ref{V892_tau_long}) \\ 
        GW Ori AB-C & 46 $\pm$ 5 & 55 $\pm$ 6 & 150 $\pm$ 7 & 137 $\pm$ 2 & 282 $\pm$ 9 & 1 $\pm$ 1 & 102 $\pm$ 9 & (\ref{gw_ori_ab_czekala_2017}) \\ 
        99 Her$^\mathrm{a}$ & 80 $\pm$ 6 & 21 $\pm$ 6 & 39 $\pm$ 2 & 45 $\pm$ 5 & 221 $\pm$ 2 & 72 $\pm$ 10 & 41 $\pm$ 2 & (\ref{99_her_kennedy}) \\ 
            SR 24N & 37 $\pm$ 6 & 96 $\pm$ 8 & 132 $\pm$ 6 & 121 $\pm$ 7 & 252 $\pm$ 4 & 297 $\pm$ 5 & 72 $\pm$ 4 & (\ref{sr_24n_andrews}), (\ref{sr_24n_fernandez}), (\ref{sr_24n_schaefer}) \\ 
        GG Tau Aa-Ab & 26 $\pm$ 6 & 80 $\pm$ 3 & 132.5 $\pm$ 2 & 143 $\pm$ 1 & 313 $\pm$ 10 & 277 $\pm$ 0.2 & 133 $\pm$ 10 & (\ref{gg_tau_a_andrews}), (\ref{gg_tau_a_difolco}), (\ref{gg_tau_a_dutrey}), (\ref{gg_tau_a_tang}), (\ref{gg_tau_a_cazzoletti})\\ 
        HD 142527 & 90 $\pm$ 2 & 159 $\pm$ 3 & 126 $\pm$ 2 & 38.2 $\pm$ 1.4 & 142 $\pm$ 6 & 162 $\pm$ 1.4 & 322 $\pm$ 6 & (\ref{hd142527_balmer}) \\ \hline
        IRS 43 & >40  & - & <30  & 70 $\pm$ 10 & - & 90 $\pm$ 5 & -  & (\ref{irs43_brinch}) \\ 
        V4046 Sgr & <2.3$^\mathrm{b}$ & - & 33.5 $\pm$ 1.4 & - & - & 256 $\pm$ 1 & - & (\ref{v4046_stempels}), (\ref{v4046_rosenfeld}), (\ref{v4046_kastner_et_al}), (\ref{v4046_kastner}) \\ 
        CoRoT 2239 & <5 & - & 85.1 $\pm$ 0.1 & 81 $\pm$ 5 & -  & -  & - & (\ref{corot_gillen_2014}), (\ref{corot_terquem}), (\ref{corot_gillen_2017})\\ 
        UZ Tau E & <2.7$^\mathrm{b}$ & - & 56.1 $\pm$ 5.7 & 56.15 $\pm$ 1.5 &  - & 269.6 $\pm$ 0.5 & - & (\ref{uz_tau_simon_2000}), (\ref{uz_tau_prato_2002}), (\ref{uz_tau_jensen_2007}) \\ 
        DQ Tau & <2.7$^\mathrm{b}$ & - & 158.2 $\pm$ 2.8 & 160 $\pm$ 3 & - & 4.2 $\pm$ 0.5 & - & (\ref{dq_tau_czekala_2016}) \\ 
        R CrA & >10 & - & 70 $\pm$ 15 & 35 $\pm$ 10 & - & 180 $\pm$ 10 & - & (\ref{r_cra}) \\ \hline
    \end{tabular}\\
    \caption{The mutual inclinations of observed circumbinary discs, with the necessary orbital parameters. The ambiguous values of longitude of ascending node ($\Omega_\mathrm{*,amb}$) are the reported value ($\Omega_\mathrm{*} + 180\degr$). $\theta_1$ are the reported mutual inclinations, $\theta_2$ are the mutual inclinations calculated using $\Omega_\mathrm{*,amb}$. The upper section of the table are the objects for which all the necessary orbital parameters required to calculate the mutual inclination have been reported. The lower section of the table are objects for which there has been an inferred, constrained value of mutual inclination reported but for which the full set of orbital parameter has not been reported.} 
    \label{tab:mutual_incs}
    \flushleft
    \textbf{Notes:} $^\mathrm{a}$ = Debris disc, $^\mathrm{b}$ = Mutual inclination inferred from hierarchical bayesian model \citep{czekala_degree_2019} \\
    \textbf{References:} (\refstepcounter{foo}\thefoo\label{ref1}) = \citet{czekala_degree_2019}; (\refstepcounter{foo}\thefoo\label{ref2}) = \citet{andrews_truncated_2010}; (\refstepcounter{foo}\thefoo\label{ref3}) = \citet{czekala_coplanar_2021}; (\refstepcounter{foo}\thefoo\label{ak_sco_anthonoiz}) = \citet{anthonioz_vltipionier_2015}; (\refstepcounter{foo}\thefoo\label{ak_sco_czekala}) = \citet{czekala_disk-based_2015}; (\refstepcounter{foo}\thefoo\label{hd_kennedy_2015}) = \citet{kennedy_nature_2015}; \refstepcounter{foo}\thefoo\label{alpha_kennedy_2012}) = \citet{kennedy_coplanar_2012}; (\refstepcounter{foo}\thefoo\label{hd_98800b_kennedy_2019}) = \citet{kennedy_circumbinary_2019} ;(\refstepcounter{foo}\thefoo\label{hd_98800b_zuniga-fernandez_2021}) = \citet{zuniga-fernandez_hd_2021}; (\refstepcounter{foo}\thefoo\label{gw_ori_ab_czekala_2017}) = \citet{czekala_architecture_2017}; (\refstepcounter{foo}\thefoo\label{hd_200775_monnier}) = \citet{monnier_few_2006}; (\refstepcounter{foo}\thefoo\label{hd_200775_okamoto_2009}) = \citet{okamoto_direct_2009}; (\refstepcounter{foo}\thefoo\label{hd_200775_benisty}) = \citet{benisty_enhanced_2013}; (\refstepcounter{foo}\thefoo\label{V892_tau_long}) = \citet{long_architecture_2021}; (\refstepcounter{foo}\thefoo\label{99_her_kennedy}) = \citet{kennedy_99_2012}; (\refstepcounter{foo}\thefoo\label{sr_24n_andrews}) = \citet{andrews_circumstellar_2005}; (\refstepcounter{foo}\thefoo\label{sr_24n_fernandez}) = \citet{fernandez-lopez_triple_misaligned_system_2017ApJ...845...10F}; (\refstepcounter{foo}\thefoo\label{sr_24n_schaefer}) = \citet{schaefer_orbital_2018}; (\refstepcounter{foo}\thefoo\label{gg_tau_a_andrews}) = \citet{andrews_resolved_2014}; (\refstepcounter{foo}\thefoo\label{gg_tau_a_difolco}) = \citet{di_folco_gg_2014}; (\refstepcounter{foo}\thefoo\label{gg_tau_a_dutrey}) = \citet{dutrey_gg_2016}; (\refstepcounter{foo}\thefoo\label{gg_tau_a_tang}) = \citet{tang_mapping_2016}; (\refstepcounter{foo}\thefoo\label{gg_tau_a_cazzoletti}) = \citet{cazzoletti_testing_2017}; (\refstepcounter{foo}\thefoo\label{hd142527_balmer}) = \citet{balmer_improved_2022}; (\refstepcounter{foo}\thefoo\label{irs43_brinch}) = \citet{brinch_IRS_43_2016ApJ...830L..16B}; (\refstepcounter{foo}\thefoo\label{v4046_stempels}) = \citet{stempels_close_2004}; (\refstepcounter{foo}\thefoo\label{v4046_rosenfeld}) = \citet{rosenfeld_disk-based_2012}; (\refstepcounter{foo}\thefoo\label{v4046_kastner_et_al}) = \citet{kastner_subarcsecond_2018}; (\refstepcounter{foo}\thefoo\label{v4046_kastner}) = \citet{kastner_candidate_2018}; (\refstepcounter{foo}\thefoo\label{corot_gillen_2014}) = \citet{gillen_corot_2014}; (\refstepcounter{foo}\thefoo\label{corot_terquem}) = \citet{terquem_circumbinary_2015}; (\refstepcounter{foo}\thefoo\label{corot_gillen_2017}) = \citet{gillen_corot_2017}; (\refstepcounter{foo}\thefoo\label{uz_tau_simon_2000}) = \citet{simon_uz_tau_2000}; (\refstepcounter{foo}\thefoo\label{uz_tau_prato_2002}) = \citet{prato_component_2002}; (\refstepcounter{foo}\thefoo\label{uz_tau_jensen_2007}) = \cite{jensen_periodic_2007}; (\refstepcounter{foo}\thefoo\label{dq_tau_czekala_2016}) = \citet{czekala_disk-based_2016}; (\refstepcounter{foo}\thefoo\label{r_cra}) = \citet{mesa_exploring_2019}
\end{table*}    

The mean mutual inclination angle of CB discs and binary orbits from the simulated systems is $\langle \theta_\mathrm{sim} \rangle = 33.6 \degr$. For the literature values it is $\langle \theta_\mathrm{obs} \rangle = 37.4 \degr$ when used the reported observational values, and when using two values for the ambiguity in the geometry of the system it is $\langle \theta_\mathrm{amb} \rangle = 70.5 \degr$. The literature value of mean mutual inclination angle is similar to the value we find in the simulation; we find a preference for alignment between discs and orbits. We perform a Kolmogorov–Smirnov test to test whether the underlying distributions of mutual inclination between the observations and simulations differ; we find no evidence at any reasonable significance level to reject the null hypothesis (p-value of $0.561$) that they do not differ. When using two values for each disc we find a distribution of mutual inclinations that is closer to a uniform distribution with a very slight preference for alignment. The value of $\langle \theta_\mathrm{obs} \rangle = 37.4 \degr$ for observed CB discs is close to the peak value of the distribution of mutual inclination of outer objects around binaries that \citet{borkovits_comprehensive_2016} find in a set of \textit{Kepler} triples. Discs that are inclined with their binary that undergo fragmentation could result in a misaligned triple system, similarly for planets forming in misaligned discs. Since the observed frequency of circumbinary planets is similar to the observed frequency of planets around single stars, and circumbinary planets with orbits that are inclined to the binary's orbital plane are more difficult to detect, this may indicate that planets around binaries are more common than those around single stars \citep{armstrong_abundance_2014}. 

\section{Conclusions}
\label{sec:conc}

We have analysed the circumbinary discs formed in a radiation hydrodynamical simulation of star cluster formation and presented their statistics. We have investigated the effect of high-order multiplicity on the properties of the circumbinary discs, and how often we might expect a circumbinary disc to form around a binary given its separation.  We have also examined how the geometries of the systems compare to observed systems.

We summarise our findings here:

\begin{enumerate}
    \item Given a sample of binaries (inclusive of those in hierarchical systems) with orbital semi-major axes less than 100~au, we would expect around $35$ per cent of the binaries to host a CB disc at young ages ($\lesssim 10^5$~years).
    \item Close binaries ($a \lesssim 3$ au) without bound companions (i.e. pure binaries) are more likely to host a CB disc than close binaries in hierarchial systems. There is a multiplicity effect.
    \item CB discs in hierarchical systems tend to be a bit less massive than those in pure binaries, and the largest discs tend to be smaller than the largest CB discs in pure binaries.
    \item The size of the CB disc scales linearly with the binary semi-major axis, $a$. The median characteristic radius of a CB disc is $\approx 5-6\ a$.
    \item CB discs of binaries with $a\gtrsim 10$ au tend to be well aligned with the binary's orbit.  In addition to this, the binaries in these well-aligned systems tend to have low eccentricity.  For binaries with small separations ($a \lesssim 3$~au) that have CB discs, the discs tend to be more randomly orientated.  This may be partially due to the finite resolution of the simulations (gas is not modelled within 0.5~au of each protostars), and the extreme youth of the systems.  
    \item The mutual inclinations of observed binaries and their CB discs are in good agreement with the mutual inclination angles found in the calculation. The underlying distributions of mutual inclination of observed and simulated CB disc do not differ according to a Kolmogorov-Smirnov test (p-value = 0.561). Mutual inclinations from the simulation have a mean value of $\langle \theta_\mathrm{sim} \rangle = 33.6 \degr$, comparatively the mean value of reported disc-orbit mutual inclinations is $\langle \theta_\mathrm{obs} \rangle = 37.4 \degr$. When assuming a $180\degr$ ambiguity for $\Omega_\mathrm{*}$ for observed binaries and CB discs the mean value of $\theta$ is $\langle \theta_\mathrm{obs,amb} \rangle = 70.5\degr$. This good agreement is obtained despite the calculation excluding some physical processes such as magnetic fields and outflows implying that such processes may not play a large role in setting the properties of protostellar discs \citep{bate_2018_10.1093/mnras/sty169,elsender_metallicity_2021}
\end{enumerate}

We hope this paper motivates an increased effort in the detection and cataloguing of circumbinary discs. We have presented evidence from simulations that circumbinary discs may be more common than what is currently being observed, especially at young ages. 

\section*{Acknowledgements}

The authors thank the anonymous referee and Ian Czekala for their comments that led to improvements to the paper. DE thanks Tim Naylor 
for useful discussions, and William O. Balmer for communications regarding HD 142527 B. DE and BSL are funded by Science and Technology Facilities Council (STFC) studentships (ST/V506679/1). DE and MRB thank the Kavli Institute of Theoretical Physics (KITP) at the University of California, Santa Barabara, the organisers and participants of binary22 at KITP.  
This research was supported in part by the National Science Foundation under Grant No. NSF PHY-1748958. This work was supported by the European Research Council under the European Commission’s Seventh Framework Programme (FP7/2007-2013 grant agreement no. 339248). The calculations discussed in this paper were performed on the University of Exeter Supercomputer, Isca, and on the DiRAC Complexity system, operated by the University of Leicester IT Services, which forms part of the STFC DiRAC HPC Facility (www.dirac.ac.uk). The latter equipment is funded by BIS National E-Infrastructure capital grant ST/K000373/1 and STFC DiRAC Operations grant ST/K0003259/1. DiRAC is part of the National e-Infrastructure.

\section*{Data Availability}

\textbf{SPH output files.} The data set consisting of the output from the radiation hydrodynamical calculation of \citet{bate_metallicity_2019MNRAS.484.2341B} that is analysed in this paper is available from the University of Exeter's Open Research Exeter (ORE) repository and can be accessed via the handle:  \url{http://hdl.handle.net/10871/3599}.

\textbf{Figure data.} The data used to create Figs \ref{fig:cb_disc_lifetime}-\ref{fig:dsratio_orient_ecc}, \ref{fig:theta} are available as supplementary material. The data needed to create Figs \ref{fig:cb_disc_lifetime}, \ref{fig:cb_disc_mass_fraction}-\ref{fig:dsratio_orient_ecc}, and \ref{fig:theta} can be found in the files titled `pair\_A\_B.txt' where the `A' and `B' are the sink numbers of the pairs as they form in the simulation. The `pair\_A\_B.txt' files have 9 columns: order of the system, mutual inclination, eccentricity, binary semi-major axis, age of the system, CB disc mass, total disc mass (primary, secondary, CB), CB disc radius, CB disc-to-binary mass ratio. The data needed to make Figs \ref{fig:cbd_fraction} and \ref{fig:binstances} are contained in `disc\_freq.txt'. This file has 3 columns: occurrence of disc (1 or 0), binary separation, order of system. It contains all instances of binaries.



\bibliographystyle{mnras}
\bibliography{refs} 

\begin{thebibliography}{}
\makeatletter
\relax
\def\mn@urlcharsother{\let\do\@makeother \do\$\do\&\do\#\do\^\do\_\do\%\do\~}
\def\mn@doi{\begingroup\mn@urlcharsother \@ifnextchar [ {\mn@doi@}
  {\mn@doi@[]}}
\def\mn@doi@[#1]#2{\def\@tempa{#1}\ifx\@tempa\@empty \href
  {http://dx.doi.org/#2} {doi:#2}\else \href {http://dx.doi.org/#2} {#1}\fi
  \endgroup}
\def\mn@eprint#1#2{\mn@eprint@#1:#2::\@nil}
\def\mn@eprint@arXiv#1{\href {http://arxiv.org/abs/#1} {{\tt arXiv:#1}}}
\def\mn@eprint@dblp#1{\href {http://dblp.uni-trier.de/rec/bibtex/#1.xml}
  {dblp:#1}}
\def\mn@eprint@#1:#2:#3:#4\@nil{\def\@tempa {#1}\def\@tempb {#2}\def\@tempc
  {#3}\ifx \@tempc \@empty \let \@tempc \@tempb \let \@tempb \@tempa \fi \ifx
  \@tempb \@empty \def\@tempb {arXiv}\fi \@ifundefined
  {mn@eprint@\@tempb}{\@tempb:\@tempc}{\expandafter \expandafter \csname
  mn@eprint@\@tempb\endcsname \expandafter{\@tempc}}}

\bibitem[\protect\citeauthoryear{{Abt} \& {Levy}}{{Abt} \&
  {Levy}}{1976}]{abt_levy_1976_multiplicity_1976ApJS...30..273A}
{Abt} H.~A.,  {Levy} S.~G.,  1976, \mn@doi [\apjs] {10.1086/190363}, \href
  {https://ui.adsabs.harvard.edu/abs/1976ApJS...30..273A} {30, 273}

\bibitem[\protect\citeauthoryear{{Akeson}, {Jensen}, {Carpenter}, {Ricci},
  {Laos}, {Nogueira}  \& {Suen-Lewis}}{{Akeson}
  et~al.}{2019}]{akeson_binaries_in_taurus_2019ApJ...872..158A}
{Akeson} R.~L.,  {Jensen} E. L.~N.,  {Carpenter} J.,  {Ricci} L.,  {Laos} S.,
  {Nogueira} N.~F.,   {Suen-Lewis} E.~M.,  2019, \mn@doi [\apj]
  {10.3847/1538-4357/aaff6a}, \href
  {https://ui.adsabs.harvard.edu/abs/2019ApJ...872..158A} {872, 158}

\bibitem[\protect\citeauthoryear{Aly \& Lodato}{Aly \&
  Lodato}{2020}]{aly_efficient_2020}
Aly H.,  Lodato G.,  2020, \mn@doi [\mnras] {10.1093/mnras/stz3633}, 492, 3306

\bibitem[\protect\citeauthoryear{Andrews \& Williams}{Andrews \&
  Williams}{2005}]{andrews_circumstellar_2005}
Andrews S.~M.,  Williams J.~P.,  2005, \mn@doi [\apj] {10.1086/432712}, 631,
  1134

\bibitem[\protect\citeauthoryear{Andrews, Czekala, Wilner, Espaillat, Dullemond
   \& Hughes}{Andrews et~al.}{2010}]{andrews_truncated_2010}
Andrews S.~M.,  Czekala I.,  Wilner D.~J.,  Espaillat C.,  Dullemond C.~P.,
  Hughes A.~M.,  2010, \mn@doi [\apj] {10.1088/0004-637X/710/1/462}, 710, 462

\bibitem[\protect\citeauthoryear{Andrews et~al.,}{Andrews
  et~al.}{2014}]{andrews_resolved_2014}
Andrews S.~M.,  et~al., 2014, \mn@doi [\apj] {10.1088/0004-637X/787/2/148},
  787, 148

\bibitem[\protect\citeauthoryear{Anthonioz et~al.,}{Anthonioz
  et~al.}{2015}]{anthonioz_vltipionier_2015}
Anthonioz F.,  et~al., 2015, \mn@doi [\aap] {10.1051/0004-6361/201424520}, 574,
  A41

\bibitem[\protect\citeauthoryear{Armstrong, Osborn, Brown, Faedi, Gómez
  Maqueo~Chew, Martin, Pollacco  \& Udry}{Armstrong
  et~al.}{2014}]{armstrong_abundance_2014}
Armstrong D.~J.,  Osborn H.~P.,  Brown D. J.~A.,  Faedi F.,  Gómez Maqueo~Chew
  Y.,  Martin D.~V.,  Pollacco D.,   Udry S.,  2014, \mn@doi [\mnras]
  {10.1093/mnras/stu1570}, 444, 1873

\bibitem[\protect\citeauthoryear{{Artymowicz} \& {Lubow}}{{Artymowicz} \&
  {Lubow}}{1994}]{artymowicz_lubow_binary_disc_interaction_1994ApJ...421..651A}
{Artymowicz} P.,  {Lubow} S.~H.,  1994, \mn@doi [\apj] {10.1086/173679}, \href
  {https://ui.adsabs.harvard.edu/abs/1994ApJ...421..651A} {421, 651}

\bibitem[\protect\citeauthoryear{{Artymowicz}, {Clarke}, {Lubow}  \&
  {Pringle}}{{Artymowicz} et~al.}{1991}]{artymowicz_orbital_evolution_1991}
{Artymowicz} P.,  {Clarke} C.~J.,  {Lubow} S.~H.,   {Pringle} J.~E.,  1991,
  \mn@doi [\apjl] {10.1086/185971}, \href
  {https://ui.adsabs.harvard.edu/abs/1991ApJ...370L..35A} {370, L35}

\bibitem[\protect\citeauthoryear{Balmer et~al.,}{Balmer
  et~al.}{2022}]{balmer_improved_2022}
Balmer W.~O.,  et~al., 2022, \mn@doi [\aj] {10.3847/1538-3881/ac73f4}, 164, 29

\bibitem[\protect\citeauthoryear{Bate}{Bate}{1998}]{bate_collapse_1998}
Bate M.~R.,  1998, \mn@doi [\apj] {10.1086/311719}, 508, L95

\bibitem[\protect\citeauthoryear{Bate}{Bate}{2012}]{bate_2012_10.1111/j.1365-2966.2011.19955.x}
Bate M.~R.,  2012, \mn@doi [MNRAS] {10.1111/j.1365-2966.2011.19955.x}, 419,
  3115

\bibitem[\protect\citeauthoryear{{Bate}}{{Bate}}{2014}]{bate_metallicity_2014MNRAS.442..285B}
{Bate} M.~R.,  2014, \mn@doi [\mnras] {10.1093/mnras/stu795}, \href
  {https://ui.adsabs.harvard.edu/abs/2014MNRAS.442..285B} {442, 285}

\bibitem[\protect\citeauthoryear{Bate}{Bate}{2018}]{bate_2018_10.1093/mnras/sty169}
Bate M.~R.,  2018, \mn@doi [\mnras] {10.1093/mnras/sty169}, 475, 5618

\bibitem[\protect\citeauthoryear{{Bate}}{{Bate}}{2019}]{bate_metallicity_2019MNRAS.484.2341B}
{Bate} M.~R.,  2019, \mn@doi [\mnras] {10.1093/mnras/stz103}, \href
  {https://ui.adsabs.harvard.edu/abs/2019MNRAS.484.2341B} {484, 2341}

\bibitem[\protect\citeauthoryear{{Bate} \& {Burkert}}{{Bate} \&
  {Burkert}}{1997}]{bate_resolution_1997MNRAS.288.1060B}
{Bate} M.~R.,  {Burkert} A.,  1997, \mn@doi [\mnras]
  {10.1093/mnras/288.4.1060}, \href
  {https://ui.adsabs.harvard.edu/abs/1997MNRAS.288.1060B} {288, 1060}

\bibitem[\protect\citeauthoryear{Bate \& Keto}{Bate \&
  Keto}{2015}]{bate_keto_10.1093/mnras/stv451}
Bate M.~R.,  Keto E.~R.,  2015, \mn@doi [MNRAS] {10.1093/mnras/stv451}, 449,
  2643

\bibitem[\protect\citeauthoryear{{Bate}, {Bonnell}  \& {Price}}{{Bate}
  et~al.}{1995}]{bate_sphng_1995MNRAS.277..362B}
{Bate} M.~R.,  {Bonnell} I.~A.,   {Price} N.~M.,  1995, \mn@doi [MNRAS]
  {10.1093/mnras/277.2.362}, \href
  {https://ui.adsabs.harvard.edu/abs/1995MNRAS.277..362B} {277, 362}

\bibitem[\protect\citeauthoryear{Bate, Bonnell  \& Bromm}{Bate
  et~al.}{2002}]{bate_formation_2002}
Bate M.~R.,  Bonnell I.~A.,   Bromm V.,  2002, \mn@doi [\mnras]
  {10.1046/j.1365-8711.2002.05775.x}, 336, 705

\bibitem[\protect\citeauthoryear{{Bate}, {Bonnell}  \& {Bromm}}{{Bate}
  et~al.}{2003}]{bate_cluster_formation_2003MNRAS.339..577B}
{Bate} M.~R.,  {Bonnell} I.~A.,   {Bromm} V.,  2003, \mn@doi [\mnras]
  {10.1046/j.1365-8711.2003.06210.x}, \href
  {https://ui.adsabs.harvard.edu/abs/2003MNRAS.339..577B} {339, 577}

\bibitem[\protect\citeauthoryear{{Bate}, {Lodato}  \& {Pringle}}{{Bate}
  et~al.}{2010}]{bate_chaotic_star_formation_2010MNRAS.401.1505B}
{Bate} M.~R.,  {Lodato} G.,   {Pringle} J.~E.,  2010, \mn@doi [MNRAS]
  {10.1111/j.1365-2966.2009.15773.x}, \href
  {https://ui.adsabs.harvard.edu/abs/2010MNRAS.401.1505B} {401, 1505}

\bibitem[\protect\citeauthoryear{Belinski, Burlak, Dodin, Emelyanov,
  Ikonnikova, Lamzin, Safonov  \& Tatarnikov}{Belinski
  et~al.}{2022}]{belinski_orbital_2022}
Belinski A.,  Burlak M.,  Dodin A.,  Emelyanov N.,  Ikonnikova N.,  Lamzin S.,
  Safonov B.,   Tatarnikov A.,  2022, \mn@doi [\mnras]
  {10.1093/mnras/stac1798}, 515, 796

\bibitem[\protect\citeauthoryear{Benisty et~al.,}{Benisty
  et~al.}{2013}]{benisty_enhanced_2013}
Benisty M.,  et~al., 2013, \mn@doi [\aap] {10.1051/0004-6361/201219893}, 555,
  A113

\bibitem[\protect\citeauthoryear{Benisty et~al.,}{Benisty
  et~al.}{2017}]{benisty_shadows_2017}
Benisty M.,  et~al., 2017, \mn@doi [\aap] {10.1051/0004-6361/201629798}, 597,
  A42

\bibitem[\protect\citeauthoryear{{Benz}}{{Benz}}{1990}]{benz_review_1990nmns.work..269B}
{Benz} W.,  1990, in {Buchler} J.~R.,  ed., Numerical Modelling of Nonlinear
  Stellar Pulsations Problems and Prospects. p.~269

\bibitem[\protect\citeauthoryear{{Benz}, {Bowers}, {Cameron}  \&
  {Press}}{{Benz} et~al.}{1990}]{benz_1990ApJ...348..647B}
{Benz} W.,  {Bowers} R.~L.,  {Cameron} A.~G.~W.,   {Press} W.~H.~.,  1990,
  \mn@doi [ApJ] {10.1086/168273}, \href
  {https://ui.adsabs.harvard.edu/abs/1990ApJ...348..647B} {348, 647}

\bibitem[\protect\citeauthoryear{Bi et~al.,}{Bi et~al.}{2020}]{bi_gw_2020}
Bi J.,  et~al., 2020, \mn@doi [The Astrophysical Journal]
  {10.3847/2041-8213/ab8eb4}, 895, L18

\bibitem[\protect\citeauthoryear{Biller et~al.,}{Biller
  et~al.}{2012}]{biller_likely_2012}
Biller B.,  et~al., 2012, \mn@doi [The Astrophysical Journal]
  {10.1088/2041-8205/753/2/L38}, 753, L38

\bibitem[\protect\citeauthoryear{Boehler, Weaver, Isella, Ricci, Grady,
  Carpenter  \& Perez}{Boehler et~al.}{2017}]{boehler_close-up_2017}
Boehler Y.,  Weaver E.,  Isella A.,  Ricci L.,  Grady C.,  Carpenter J.,
  Perez L.,  2017, \mn@doi [The Astrophysical Journal]
  {10.3847/1538-4357/aa696c}, 840, 60

\bibitem[\protect\citeauthoryear{Bonnell \& Bate}{Bonnell \&
  Bate}{1994}]{bonnell_formation_1994}
Bonnell I.~A.,  Bate M.~R.,  1994, \mn@doi [\mnras] {10.1093/mnras/271.4.999},
  271, 999

\bibitem[\protect\citeauthoryear{Borkovits, Hajdu, Sztakovics, Rappaport,
  Levine, Bíró  \& Klagyivik}{Borkovits
  et~al.}{2016}]{borkovits_comprehensive_2016}
Borkovits T.,  Hajdu T.,  Sztakovics J.,  Rappaport S.,  Levine A.,  Bíró
  I.~B.,   Klagyivik P.,  2016, \mn@doi [\mnras] {10.1093/mnras/stv2530}, 455,
  4136

\bibitem[\protect\citeauthoryear{Boss}{Boss}{1986}]{boss_protostellar_1986}
Boss A.~P.,  1986, \mn@doi [\apjs] {10.1086/191150}, 62, 519

\bibitem[\protect\citeauthoryear{Boss \& Bodenheimer}{Boss \&
  Bodenheimer}{1979}]{boss_fragmentation_1979}
Boss A.~P.,  Bodenheimer P.,  1979, \mn@doi [\apj] {10.1086/157497}, 234, 289

\bibitem[\protect\citeauthoryear{{Boss}, {Fisher}, {Klein}  \& {McKee}}{{Boss}
  et~al.}{2000}]{boss_jeans_2000ApJ...528..325B}
{Boss} A.~P.,  {Fisher} R.~T.,  {Klein} R.~I.,   {McKee} C.~F.,  2000, \mn@doi
  [\apj] {10.1086/308160}, \href
  {https://ui.adsabs.harvard.edu/abs/2000ApJ...528..325B} {528, 325}

\bibitem[\protect\citeauthoryear{Brinch, Jørgensen, Hogerheijde, Nelson  \&
  Gressel}{Brinch et~al.}{2016a}]{brinch_misaligned_2016}
Brinch C.,  Jørgensen J.~K.,  Hogerheijde M.~R.,  Nelson R.~P.,   Gressel O.,
  2016a, \mn@doi [\apj] {10.3847/2041-8205/830/1/L16}, 830, L16

\bibitem[\protect\citeauthoryear{{Brinch}, {J{\o}rgensen}, {Hogerheijde},
  {Nelson}  \& {Gressel}}{{Brinch}
  et~al.}{2016b}]{brinch_IRS_43_2016ApJ...830L..16B}
{Brinch} C.,  {J{\o}rgensen} J.~K.,  {Hogerheijde} M.~R.,  {Nelson} R.~P.,
  {Gressel} O.,  2016b, \mn@doi [ApJL] {10.3847/2041-8205/830/1/L16}, \href
  {https://ui.adsabs.harvard.edu/abs/2016ApJ...830L..16B} {830, L16}

\bibitem[\protect\citeauthoryear{Casassus et~al.,}{Casassus
  et~al.}{2018}]{casassus_inner_2018}
Casassus S.,  et~al., 2018, \mn@doi [\mnras] {10.1093/mnras/sty894}, 477, 5104

\bibitem[\protect\citeauthoryear{Cazzoletti, Ricci, Birnstiel  \&
  Lodato}{Cazzoletti et~al.}{2017}]{cazzoletti_testing_2017}
Cazzoletti P.,  Ricci L.,  Birnstiel T.,   Lodato G.,  2017, \mn@doi [\aap]
  {10.1051/0004-6361/201629721}, 599, A102

\bibitem[\protect\citeauthoryear{Chiang \& Murray-Clay}{Chiang \&
  Murray-Clay}{2004}]{chiang_circumbinary_2004}
Chiang E.~I.,  Murray-Clay R.~A.,  2004, \mn@doi [\apj] {10.1086/383522}, 607,
  913

\bibitem[\protect\citeauthoryear{Clarke}{Clarke}{2009}]{clarke_pseudo-viscous_2009}
Clarke C.~J.,  2009, \mn@doi [\mnras] {10.1111/j.1365-2966.2009.14774.x}, 396,
  1066

\bibitem[\protect\citeauthoryear{Claudi et~al.,}{Claudi
  et~al.}{2019}]{claudi_sphere_2019}
Claudi R.,  et~al., 2019, \mn@doi [Astronomy and Astrophysics]
  {10.1051/0004-6361/201833990}, 622, A96

\bibitem[\protect\citeauthoryear{Czekala, Andrews, Jensen, Stassun, Torres  \&
  Wilner}{Czekala et~al.}{2015}]{czekala_disk-based_2015}
Czekala I.,  Andrews S.~M.,  Jensen E. L.~N.,  Stassun K.~G.,  Torres G.,
  Wilner D.~J.,  2015, \mn@doi [\apj] {10.1088/0004-637X/806/2/154}, 806, 154

\bibitem[\protect\citeauthoryear{Czekala, Andrews, Torres, Jensen, Stassun,
  Wilner  \& Latham}{Czekala et~al.}{2016}]{czekala_disk-based_2016}
Czekala I.,  Andrews S.~M.,  Torres G.,  Jensen E. L.~N.,  Stassun K.~G.,
  Wilner D.~J.,   Latham D.~W.,  2016, \mn@doi [\apj]
  {10.3847/0004-637X/818/2/156}, 818, 156

\bibitem[\protect\citeauthoryear{Czekala et~al.,}{Czekala
  et~al.}{2017}]{czekala_architecture_2017}
Czekala I.,  et~al., 2017, \mn@doi [\apj] {10.3847/1538-4357/aa9be7}, 851, 132

\bibitem[\protect\citeauthoryear{Czekala, Chiang, Andrews, Jensen, Torres,
  Wilner, Stassun  \& Macintosh}{Czekala et~al.}{2019}]{czekala_degree_2019}
Czekala I.,  Chiang E.,  Andrews S.~M.,  Jensen E. L.~N.,  Torres G.,  Wilner
  D.~J.,  Stassun K.~G.,   Macintosh B.,  2019, \mn@doi [\apj]
  {10.3847/1538-4357/ab287b}, 883, 22

\bibitem[\protect\citeauthoryear{Czekala, Ribas, Cuello, Chiang, Mac\'{i}as,
  Duch\^{e}ne, Andrews  \& Espaillat}{Czekala
  et~al.}{2021}]{czekala_coplanar_2021}
Czekala I.,  Ribas A.,  Cuello N.,  Chiang E.,  Mac\'{i}as E.,  Duch\^{e}ne G.,
   Andrews S.~M.,   Espaillat C.~C.,  2021, \mn@doi [\apj]
  {10.3847/1538-4357/abebe3}, 912, 6

\bibitem[\protect\citeauthoryear{Di~Folco et~al.,}{Di~Folco
  et~al.}{2014}]{di_folco_gg_2014}
Di~Folco E.,  et~al., 2014, \mn@doi [\aap] {10.1051/0004-6361/201423675}, 565,
  L2

\bibitem[\protect\citeauthoryear{Doyle et~al.,}{Doyle et~al.}{2011}]{Doyle1602}
Doyle L.~R.,  et~al., 2011, \mn@doi [Science] {10.1126/science.1210923}, 333,
  1602

\bibitem[\protect\citeauthoryear{{Duquennoy} \& {Mayor}}{{Duquennoy} \&
  {Mayor}}{1991}]{duquennoy_mayor_multiplicity_1991A&A...248..485D}
{Duquennoy} A.,  {Mayor} M.,  1991, \aap, \href
  {https://ui.adsabs.harvard.edu/abs/1991A&A...248..485D} {248, 485}

\bibitem[\protect\citeauthoryear{Dutrey, Guilloteau  \& Simon}{Dutrey
  et~al.}{1994}]{dutrey1994images}
Dutrey A.,  Guilloteau S.,   Simon M.,  1994, \aap, 286, 149

\bibitem[\protect\citeauthoryear{Dutrey, Di~Folco, Beck  \& Guilloteau}{Dutrey
  et~al.}{2016}]{dutrey_gg_2016}
Dutrey A.,  Di~Folco E.,  Beck T.,   Guilloteau S.,  2016, \mn@doi [\aapr]
  {10.1007/s00159-015-0091-5}, 24, 5

\bibitem[\protect\citeauthoryear{Duvert, Dutrey, Guilloteau, Menard, Schuster,
  Prato  \& Simon}{Duvert et~al.}{1998}]{duvert_disks_1998}
Duvert G.,  Dutrey A.,  Guilloteau S.,  Menard F.,  Schuster K.,  Prato L.,
  Simon M.,  1998, \aap, 332, 867

\bibitem[\protect\citeauthoryear{{Dvorak}}{{Dvorak}}{1982}]{dvorak_orbit_types_1982OAWMN.191..423D}
{Dvorak} R.,  1982, Oesterreichische Akademie Wissenschaften Mathematisch
  naturwissenschaftliche Klasse Sitzungsberichte Abteilung, \href
  {https://ui.adsabs.harvard.edu/abs/1982OAWMN.191..423D} {191, 423}

\bibitem[\protect\citeauthoryear{{Elsender} \& {Bate}}{{Elsender} \&
  {Bate}}{2021}]{elsender_metallicity_2021}
{Elsender} D.,  {Bate} M.~R.,  2021, \mn@doi [\mnras] {10.1093/mnras/stab2901},
  \href {https://ui.adsabs.harvard.edu/abs/2021MNRAS.508.5279E} {508, 5279}

\bibitem[\protect\citeauthoryear{Facchini, Lodato  \& Price}{Facchini
  et~al.}{2013}]{facchini_wave-like_2013}
Facchini S.,  Lodato G.,   Price D.~J.,  2013, \mn@doi [\mnras]
  {10.1093/mnras/stt877}, 433, 2142

\bibitem[\protect\citeauthoryear{{Fedele} et~al.,}{{Fedele}
  et~al.}{2017}]{fedele_rings_and_gaps_in_ppd_2017A&A...600A..72F}
{Fedele} D.,  et~al., 2017, \mn@doi [A\&A] {10.1051/0004-6361/201629860}, \href
  {https://ui.adsabs.harvard.edu/abs/2017A&A...600A..72F} {600, A72}

\bibitem[\protect\citeauthoryear{{Fern{\'a}ndez-L{\'o}pez}, {Zapata}  \&
  {Gabbasov}}{{Fern{\'a}ndez-L{\'o}pez}
  et~al.}{2017}]{fernandez-lopez_triple_misaligned_system_2017ApJ...845...10F}
{Fern{\'a}ndez-L{\'o}pez} M.,  {Zapata} L.~A.,   {Gabbasov} R.,  2017, \mn@doi
  [ApJ] {10.3847/1538-4357/aa7d51}, \href
  {https://ui.adsabs.harvard.edu/abs/2017ApJ...845...10F} {845, 10}

\bibitem[\protect\citeauthoryear{Fisher}{Fisher}{2004}]{fisher_turbulent_2004}
Fisher R.~T.,  2004, \mn@doi [\apj] {10.1086/380111}, 600, 769

\bibitem[\protect\citeauthoryear{Fu, Lubow  \& Martin}{Fu
  et~al.}{2015}]{fu_kozai-lidov_2015}
Fu W.,  Lubow S.~H.,   Martin R.~G.,  2015, \mn@doi [\apj]
  {10.1088/0004-637X/807/1/75}, 807, 75

\bibitem[\protect\citeauthoryear{Ghez, Neugebauer  \& Matthews}{Ghez
  et~al.}{1993}]{ghez1993multiplicity}
Ghez A.,  Neugebauer G.,   Matthews K.,  1993, Astronomical Journal, 106, 2005

\bibitem[\protect\citeauthoryear{Gillen et~al.,}{Gillen
  et~al.}{2014}]{gillen_corot_2014}
Gillen E.,  et~al., 2014, \mn@doi [\aap] {10.1051/0004-6361/201322493}, 562,
  A50

\bibitem[\protect\citeauthoryear{Gillen et~al.,}{Gillen
  et~al.}{2017}]{gillen_corot_2017}
Gillen E.,  et~al., 2017, \mn@doi [\aap] {10.1051/0004-6361/201628483}, 599,
  A27

\bibitem[\protect\citeauthoryear{{Glover} \& {Clark}}{{Glover} \&
  {Clark}}{2012}]{glover_metal_poor_gas_2012MNRAS.426..377G}
{Glover} S. C.~O.,  {Clark} P.~C.,  2012, \mn@doi [\mnras]
  {10.1111/j.1365-2966.2012.21737.x}, \href
  {https://ui.adsabs.harvard.edu/abs/2012MNRAS.426..377G} {426, 377}

\bibitem[\protect\citeauthoryear{{Harris}, {Andrews}, {Wilner}  \&
  {Kraus}}{{Harris}
  et~al.}{2012}]{harris_taurus_multiple_systems_2012ApJ...751..115H}
{Harris} R.~J.,  {Andrews} S.~M.,  {Wilner} D.~J.,   {Kraus} A.~L.,  2012,
  \mn@doi [\apj] {10.1088/0004-637X/751/2/115}, \href
  {https://ui.adsabs.harvard.edu/abs/2012ApJ...751..115H} {751, 115}

\bibitem[\protect\citeauthoryear{Harris et~al.,}{Harris
  et~al.}{2018}]{harris_alma_2018}
Harris R.~J.,  et~al., 2018, \mn@doi [\apj] {10.3847/1538-4357/aac6ec}, 861, 91

\bibitem[\protect\citeauthoryear{{Hsieh}, {Lai}, {Cheong}, {Ko}, {Li}  \&
  {Murillo}}{{Hsieh} et~al.}{2020}]{hsieh_vla1623a_2020ApJ...894...23H}
{Hsieh} C.-H.,  {Lai} S.-P.,  {Cheong} P.-I.,  {Ko} C.-L.,  {Li} Z.-Y.,
  {Murillo} N.~M.,  2020, \mn@doi [\apj] {10.3847/1538-4357/ab7b69}, \href
  {https://ui.adsabs.harvard.edu/abs/2020ApJ...894...23H} {894, 23}

\bibitem[\protect\citeauthoryear{{Hubber}, {Goodwin}  \& {Whitworth}}{{Hubber}
  et~al.}{2006}]{hubber_resolution_2006A&A...450..881H}
{Hubber} D.~A.,  {Goodwin} S.~P.,   {Whitworth} A.~P.,  2006, \mn@doi [\aap]
  {10.1051/0004-6361:20054100}, \href
  {https://ui.adsabs.harvard.edu/abs/2006A&A...450..881H} {450, 881}

\bibitem[\protect\citeauthoryear{Jensen \& Mathieu}{Jensen \&
  Mathieu}{1997}]{jensen_evidence_1997}
Jensen E. L.~N.,  Mathieu R.~D.,  1997, \mn@doi [\aj] {10.1086/118475}, 114,
  301

\bibitem[\protect\citeauthoryear{Jensen, Koerner  \& Mathieu}{Jensen
  et~al.}{1996a}]{jensen_high-resolution_1996}
Jensen E. L.~N.,  Koerner D.~W.,   Mathieu R.~D.,  1996a, \mn@doi [\aj]
  {10.1086/117977}, 111, 2431

\bibitem[\protect\citeauthoryear{{Jensen}, {Mathieu}  \& {Fuller}}{{Jensen}
  et~al.}{1996b}]{jensen_separation_1996ApJ...458..312J}
{Jensen} E. L.~N.,  {Mathieu} R.~D.,   {Fuller} G.~A.,  1996b, \mn@doi [\apj]
  {10.1086/176814}, \href
  {https://ui.adsabs.harvard.edu/abs/1996ApJ...458..312J} {458, 312}

\bibitem[\protect\citeauthoryear{Jensen, Dhital, Stassun, Patience, Herbst,
  Walter, Simon  \& Basri}{Jensen et~al.}{2007}]{jensen_periodic_2007}
Jensen E. L.~N.,  Dhital S.,  Stassun K.~G.,  Patience J.,  Herbst W.,  Walter
  F.~M.,  Simon M.,   Basri G.,  2007, \mn@doi [\aj] {10.1086/518408}, 134, 241

\bibitem[\protect\citeauthoryear{Kastner}{Kastner}{2018}]{kastner_candidate_2018}
Kastner J.~H.,  2018, \mn@doi [Res. Notes Am. Astron. Soc.]
  {10.3847/2515-5172/aad62c}, 2, 137

\bibitem[\protect\citeauthoryear{Kastner et~al.,}{Kastner
  et~al.}{2018}]{kastner_subarcsecond_2018}
Kastner J.~H.,  et~al., 2018, \mn@doi [\apj] {10.3847/1538-4357/aacff7}, 863,
  106

\bibitem[\protect\citeauthoryear{Kawabe, Ishiguro, Omodaka, Kitamura  \&
  Miyama}{Kawabe et~al.}{1993}]{kawabe_discovery_1993}
Kawabe R.,  Ishiguro M.,  Omodaka T.,  Kitamura Y.,   Miyama S.~M.,  1993,
  \mn@doi [\apj] {10.1086/186744}, 404, L63

\bibitem[\protect\citeauthoryear{Kennedy}{Kennedy}{2015}]{kennedy_nature_2015}
Kennedy G.~M.,  2015, \mn@doi [\mnras] {10.1093/mnrasl/slu190}, 447, L75

\bibitem[\protect\citeauthoryear{Kennedy et~al.,}{Kennedy
  et~al.}{2012a}]{kennedy_99_2012}
Kennedy G.~M.,  et~al., 2012a, \mn@doi [\mnras]
  {10.1111/j.1365-2966.2012.20448.x}, 421, 2264

\bibitem[\protect\citeauthoryear{Kennedy, Wyatt, Sibthorpe, Phillips, Matthews
  \& Greaves}{Kennedy et~al.}{2012b}]{kennedy_coplanar_2012}
Kennedy G.~M.,  Wyatt M.~C.,  Sibthorpe B.,  Phillips N.~M.,  Matthews B.~C.,
  Greaves J.~S.,  2012b, \mn@doi [\mnras] {10.1111/j.1365-2966.2012.21865.x},
  426, 2115

\bibitem[\protect\citeauthoryear{Kennedy et~al.,}{Kennedy
  et~al.}{2019}]{kennedy_circumbinary_2019}
Kennedy G.~M.,  et~al., 2019, \mn@doi [Nature Astronomy]
  {10.1038/s41550-018-0667-x}, 3, 230

\bibitem[\protect\citeauthoryear{Kenworthy et~al.,}{Kenworthy
  et~al.}{2022}]{kenworthy_eclipse_2022}
Kenworthy M.~A.,  et~al., 2022, \mn@doi [\aap] {10.1051/0004-6361/202243441},
  666, A61

\bibitem[\protect\citeauthoryear{{Kostov} et~al.,}{{Kostov}
  et~al.}{2020}]{kostov_TESS_TOI_1338_2020AJ....159..253K}
{Kostov} V.~B.,  et~al., 2020, \mn@doi [\aj] {10.3847/1538-3881/ab8a48}, \href
  {https://ui.adsabs.harvard.edu/abs/2020AJ....159..253K} {159, 253}

\bibitem[\protect\citeauthoryear{{Kostov} et~al.,}{{Kostov}
  et~al.}{2021}]{kostov_TIC_172900988_2021AJ....162..234K}
{Kostov} V.~B.,  et~al., 2021, \mn@doi [\aj] {10.3847/1538-3881/ac223a}, \href
  {https://ui.adsabs.harvard.edu/abs/2021AJ....162..234K} {162, 234}

\bibitem[\protect\citeauthoryear{Kozai}{Kozai}{1962}]{kozai_secular_1962}
Kozai Y.,  1962, \mn@doi [\aj] {10.1086/108790}, 67, 591

\bibitem[\protect\citeauthoryear{Kratter \& Matzner}{Kratter \&
  Matzner}{2006}]{kratter_fragmentation_2006}
Kratter K.~M.,  Matzner C.~D.,  2006, \mn@doi [\mnras]
  {10.1111/j.1365-2966.2006.11103.x}, 373, 1563

\bibitem[\protect\citeauthoryear{Kraus, Ireland, Hillenbrand  \&
  Martinache}{Kraus et~al.}{2012}]{kraus_role_2012}
Kraus A.~L.,  Ireland M.~J.,  Hillenbrand L.~A.,   Martinache F.,  2012,
  \mn@doi [The Astrophysical Journal] {10.1088/0004-637X/745/1/19}, 745, 19

\bibitem[\protect\citeauthoryear{Kraus et~al.,}{Kraus
  et~al.}{2020}]{kraus_triple-star_2020}
Kraus S.,  et~al., 2020, \mn@doi [Science] {10.1126/science.aba4633}, 369, 1233

\bibitem[\protect\citeauthoryear{Kuruwita \& Federrath}{Kuruwita \&
  Federrath}{2019}]{kuruwita_role_2019}
Kuruwita R.~L.,  Federrath C.,  2019, \mn@doi [\mnras] {10.1093/mnras/stz1053},
  486, 3647

\bibitem[\protect\citeauthoryear{Lacour et~al.,}{Lacour
  et~al.}{2016}]{lacour_m-dwarf_2016}
Lacour S.,  et~al., 2016, \mn@doi [Astronomy and Astrophysics]
  {10.1051/0004-6361/201527863}, 590, A90

\bibitem[\protect\citeauthoryear{{Larson}}{{Larson}}{1969}]{larson_1969MNRAS.145..271L}
{Larson} R.~B.,  1969, \mn@doi [\mnras] {10.1093/mnras/145.3.271}, \href
  {https://ui.adsabs.harvard.edu/abs/1969MNRAS.145..271L} {145, 271}

\bibitem[\protect\citeauthoryear{Lidov}{Lidov}{1962}]{lidov_evolution_1962}
Lidov M.~L.,  1962, \mn@doi [Planetary and Space Science]
  {10.1016/0032-0633(62)90129-0}, 9, 719

\bibitem[\protect\citeauthoryear{{Lin} \& {Papaloizou}}{{Lin} \&
  {Papaloizou}}{1979a}]{lin_papaloizou_tidal_torques_1979MNRAS.186..799L}
{Lin} D.~N.~C.,  {Papaloizou} J.,  1979a, \mn@doi [\mnras]
  {10.1093/mnras/186.4.799}, \href
  {https://ui.adsabs.harvard.edu/abs/1979MNRAS.186..799L} {186, 799}

\bibitem[\protect\citeauthoryear{{Lin} \& {Papaloizou}}{{Lin} \&
  {Papaloizou}}{1979b}]{lin_papaloizou_structure_of_cb_discs_1979MNRAS.188..191L}
{Lin} D.~N.~C.,  {Papaloizou} J.,  1979b, \mn@doi [\mnras]
  {10.1093/mnras/188.2.191}, \href
  {https://ui.adsabs.harvard.edu/abs/1979MNRAS.188..191L} {188, 191}

\bibitem[\protect\citeauthoryear{Long et~al.,}{Long
  et~al.}{2021}]{long_architecture_2021}
Long F.,  et~al., 2021, \mn@doi [\apj] {10.3847/1538-4357/abff53}, 915, 131

\bibitem[\protect\citeauthoryear{Lubow \& Martin}{Lubow \&
  Martin}{2018}]{lubow_linear_2018}
Lubow S.~H.,  Martin R.~G.,  2018, \mn@doi [\mnras] {10.1093/mnras/stx2643},
  473, 3733

\bibitem[\protect\citeauthoryear{Marino, Perez  \& Casassus}{Marino
  et~al.}{2015}]{marino_shadows_2015}
Marino S.,  Perez S.,   Casassus S.,  2015, \mn@doi [\apj]
  {10.1088/2041-8205/798/2/L44}, 798, L44

\bibitem[\protect\citeauthoryear{{Martin} \& {Lubow}}{{Martin} \&
  {Lubow}}{2017}]{martin_lubow_polar_alignment_i_2017ApJ...835L..28M}
{Martin} R.~G.,  {Lubow} S.~H.,  2017, \mn@doi [\apjl]
  {10.3847/2041-8213/835/2/L28}, \href
  {https://ui.adsabs.harvard.edu/abs/2017ApJ...835L..28M} {835, L28}

\bibitem[\protect\citeauthoryear{Martin \& Lubow}{Martin \&
  Lubow}{2019}]{martin_polar_2019}
Martin R.~G.,  Lubow S.~H.,  2019, \mn@doi [\mnras] {10.1093/mnras/stz2670},
  490, 1332

\bibitem[\protect\citeauthoryear{Martin \& Lubow}{Martin \&
  Lubow}{2022}]{martin_eccentric_2022}
Martin R.~G.,  Lubow S.~H.,  2022, \mn@doi [\apj] {10.3847/2041-8213/ac4974},
  925, L1

\bibitem[\protect\citeauthoryear{Martin \& Triaud}{Martin \&
  Triaud}{2014}]{martin_planets_2014}
Martin D.~V.,  Triaud A. H. M.~J.,  2014, \mn@doi [\aap]
  {10.1051/0004-6361/201323112}, 570, A91

\bibitem[\protect\citeauthoryear{Martin \& Triaud}{Martin \&
  Triaud}{2015}]{martin_circumbinary_2015}
Martin D.~V.,  Triaud A. H. M.~J.,  2015, \mn@doi [\mnras]
  {10.1093/mnras/stv121}, 449, 781

\bibitem[\protect\citeauthoryear{Martin, Nixon, Lubow, Armitage, Price, Doğan
  \& King}{Martin et~al.}{2014}]{martin_kozai-lidov_2014}
Martin R.~G.,  Nixon C.,  Lubow S.~H.,  Armitage P.~J.,  Price D.~J.,  Doğan
  S.,   King A.,  2014, \mn@doi [\apj] {10.1088/2041-8205/792/2/L33}, 792, L33

\bibitem[\protect\citeauthoryear{Mathieu, Adams, Fuller, Jensen, Koerner  \&
  Sargent}{Mathieu et~al.}{1995}]{mathieu_submillimeter_1995}
Mathieu R.~D.,  Adams F.~C.,  Fuller G.~A.,  Jensen E. L.~N.,  Koerner D.~W.,
  Sargent A.~I.,  1995, \mn@doi [\aj] {10.1086/117479}, 109, 2655

\bibitem[\protect\citeauthoryear{Mathieu, Martin  \& Magazzu}{Mathieu
  et~al.}{1996}]{mathieu_uz_1996}
Mathieu R.~D.,  Martin E.~L.,   Magazzu A.,  1996, 188, 60.05

\bibitem[\protect\citeauthoryear{Mathieu, Stassun, Basri, Jensen, Johns-Krull,
  Valenti  \& Hartmann}{Mathieu et~al.}{1997}]{mathieu_classical_1997}
Mathieu R.~D.,  Stassun K.,  Basri G.,  Jensen E. L.~N.,  Johns-Krull C.~M.,
  Valenti J.~A.,   Hartmann L.~W.,  1997, \mn@doi [\aj] {10.1086/118395}, 113,
  1841

\bibitem[\protect\citeauthoryear{Maureira, Pineda, Segura-Cox, Caselli, Testi,
  Lodato, Loinard  \& Hernández-Gómez}{Maureira
  et~al.}{2020}]{maureira_orbital_2020}
Maureira M.~J.,  Pineda J.~E.,  Segura-Cox D.~M.,  Caselli P.,  Testi L.,
  Lodato G.,  Loinard L.,   Hernández-Gómez A.,  2020, \mn@doi [\apj]
  {10.3847/1538-4357/ab960b}, 897, 59

\bibitem[\protect\citeauthoryear{McKee \& Ostriker}{McKee \&
  Ostriker}{2007}]{mckee_theory_2007}
McKee C.~F.,  Ostriker E.~C.,  2007, \mn@doi [\araa]
  {10.1146/annurev.astro.45.051806.110602}, 45, 565

\bibitem[\protect\citeauthoryear{Mesa et~al.,}{Mesa
  et~al.}{2019}]{mesa_exploring_2019}
Mesa D.,  et~al., 2019, \mn@doi [\aap] {10.1051/0004-6361/201834682}, 624, A4

\bibitem[\protect\citeauthoryear{Monnier et~al.,}{Monnier
  et~al.}{2006}]{monnier_few_2006}
Monnier J.~D.,  et~al., 2006, \mn@doi [\apj] {10.1086/505340}, 647, 444

\bibitem[\protect\citeauthoryear{Nixon, King  \& Price}{Nixon
  et~al.}{2013}]{nixon_tearing_2013}
Nixon C.,  King A.,   Price D.,  2013, \mn@doi [\mnras]
  {10.1093/mnras/stt1136}, 434, 1946

\bibitem[\protect\citeauthoryear{Okamoto et~al.,}{Okamoto
  et~al.}{2009}]{okamoto_direct_2009}
Okamoto Y.~K.,  et~al., 2009, \mn@doi [\apj] {10.1088/0004-637X/706/1/665},
  706, 665

\bibitem[\protect\citeauthoryear{{Ostriker}, {Stone}  \& {Gammie}}{{Ostriker}
  et~al.}{2001}]{Ostriker_turbulence_2001ApJ...546..980O}
{Ostriker} E.~C.,  {Stone} J.~M.,   {Gammie} C.~F.,  2001, \mn@doi [\apj]
  {10.1086/318290}, \href
  {https://ui.adsabs.harvard.edu/abs/2001ApJ...546..980O} {546, 980}

\bibitem[\protect\citeauthoryear{{Papaloizou} \& {Pringle}}{{Papaloizou} \&
  {Pringle}}{1977}]{papaloizou_pringle_tidal_torques_1977MNRAS.181..441P}
{Papaloizou} J.,  {Pringle} J.~E.,  1977, \mn@doi [\mnras]
  {10.1093/mnras/181.3.441}, \href
  {https://ui.adsabs.harvard.edu/abs/1977MNRAS.181..441P} {181, 441}

\bibitem[\protect\citeauthoryear{Prato, Simon, Mazeh, Zucker  \& McLean}{Prato
  et~al.}{2002}]{prato_component_2002}
Prato L.,  Simon M.,  Mazeh T.,  Zucker S.,   McLean I.~S.,  2002, \mn@doi
  [\apj] {10.1086/345317}, 579, L99

\bibitem[\protect\citeauthoryear{{Price} \& {Monaghan}}{{Price} \&
  {Monaghan}}{2007}]{price_2007MNRAS.374.1347P}
{Price} D.~J.,  {Monaghan} J.~J.,  2007, \mn@doi [\mnras]
  {10.1111/j.1365-2966.2006.11241.x}, \href
  {https://ui.adsabs.harvard.edu/abs/2007MNRAS.374.1347P} {374, 1347}

\bibitem[\protect\citeauthoryear{{Price} et~al.,}{{Price}
  et~al.}{2018}]{price_circumbinary_not_transitional_2018MNRAS.477.1270P}
{Price} D.~J.,  et~al., 2018, \mn@doi [MNRAS] {10.1093/mnras/sty647}, \href
  {https://ui.adsabs.harvard.edu/abs/2018MNRAS.477.1270P} {477, 1270}

\bibitem[\protect\citeauthoryear{{Raghavan} et~al.,}{{Raghavan}
  et~al.}{2010}]{raghavan_multiplicity_2010ApJS..190....1R}
{Raghavan} D.,  et~al., 2010, \mn@doi [\apjs] {10.1088/0067-0049/190/1/1},
  \href {https://ui.adsabs.harvard.edu/abs/2010ApJS..190....1R} {190, 1}

\bibitem[\protect\citeauthoryear{Rosenfeld, Andrews, Wilner  \&
  Stempels}{Rosenfeld et~al.}{2012}]{rosenfeld_disk-based_2012}
Rosenfeld K.~A.,  Andrews S.~M.,  Wilner D.~J.,   Stempels H.~C.,  2012,
  \mn@doi [\apj] {10.1088/0004-637X/759/2/119}, 759, 119

\bibitem[\protect\citeauthoryear{Sadavoy et~al.,}{Sadavoy
  et~al.}{2018}]{sadavoy_dust_2018}
Sadavoy S.~I.,  et~al., 2018, \mn@doi [\apj] {10.3847/1538-4357/aac21a}, 859,
  165

\bibitem[\protect\citeauthoryear{Schaefer, Prato  \& Simon}{Schaefer
  et~al.}{2018}]{schaefer_orbital_2018}
Schaefer G.~H.,  Prato L.,   Simon M.,  2018, \mn@doi [\aj]
  {10.3847/1538-3881/aaa59a}, 155, 109

\bibitem[\protect\citeauthoryear{{Simon}, {Dutrey}  \& {Guilloteau}}{{Simon}
  et~al.}{2000}]{simon_uz_tau_2000}
{Simon} M.,  {Dutrey} A.,   {Guilloteau} S.,  2000, \mn@doi [\apj]
  {10.1086/317838}, \href
  {https://ui.adsabs.harvard.edu/abs/2000ApJ...545.1034S} {545, 1034}

\bibitem[\protect\citeauthoryear{{Standing} et~al.,}{{Standing}
  et~al.}{2023}]{standing_first_2023}
{Standing} M.~R.,  et~al., 2023, \mn@doi [arXiv e-prints]
  {10.48550/arXiv.2301.10794}, \href
  {https://ui.adsabs.harvard.edu/abs/2023arXiv230110794S} {p. arXiv:2301.10794}

\bibitem[\protect\citeauthoryear{Stempels \& Gahm}{Stempels \&
  Gahm}{2004}]{stempels_close_2004}
Stempels H.~C.,  Gahm G.~F.,  2004, \mn@doi [\aap]
  {10.1051/0004-6361:20034502}, 421, 1159

\bibitem[\protect\citeauthoryear{Sterzik \& Tokovinin}{Sterzik \&
  Tokovinin}{2002}]{sterzik_relative_2002}
Sterzik M.~F.,  Tokovinin A.~A.,  2002, \mn@doi [\aap]
  {10.1051/0004-6361:20020105}, 384, 1030

\bibitem[\protect\citeauthoryear{{Takakuwa}, {Saito}, {Lim}, {Saigo},
  {Sridharan}  \& {Patel}}{{Takakuwa}
  et~al.}{2012}]{takakuwa_L1551_NE_2012ApJ...754...52T}
{Takakuwa} S.,  {Saito} M.,  {Lim} J.,  {Saigo} K.,  {Sridharan} T.~K.,
  {Patel} N.~A.,  2012, \mn@doi [\apj] {10.1088/0004-637X/754/1/52}, \href
  {https://ui.adsabs.harvard.edu/abs/2012ApJ...754...52T} {754, 52}

\bibitem[\protect\citeauthoryear{Tang et~al.,}{Tang
  et~al.}{2016}]{tang_mapping_2016}
Tang Y.-W.,  et~al., 2016, \mn@doi [\apj] {10.3847/0004-637X/820/1/19}, 820, 19

\bibitem[\protect\citeauthoryear{{Tazzari} et~al.,}{{Tazzari}
  et~al.}{2017}]{2017A&A...606A..88T}
{Tazzari} M.,  et~al., 2017, \mn@doi [\aap] {10.1051/0004-6361/201730890},
  \href {https://ui.adsabs.harvard.edu/abs/2017A&A...606A..88T} {606, A88}

\bibitem[\protect\citeauthoryear{Terquem, Sørensen-Clark  \& Bouvier}{Terquem
  et~al.}{2015}]{terquem_circumbinary_2015}
Terquem C.,  Sørensen-Clark P.~M.,   Bouvier J.,  2015, \mn@doi [\mnras]
  {10.1093/mnras/stv2258}, 454, 3472

\bibitem[\protect\citeauthoryear{Tobin et~al.,}{Tobin
  et~al.}{2016a}]{tobin_triple_2016}
Tobin J.,  et~al., 2016a, \mn@doi [Nature] {10.1038/nature20094}, 538, 483

\bibitem[\protect\citeauthoryear{Tobin et~al.,}{Tobin
  et~al.}{2016b}]{tobin_vla_2016}
Tobin J.~J.,  et~al., 2016b, \mn@doi [\apj] {10.3847/0004-637X/818/1/73}, 818,
  73

\bibitem[\protect\citeauthoryear{Tobin et~al.,}{Tobin
  et~al.}{2018}]{tobin_vlaalma_2018}
Tobin J.~J.,  et~al., 2018, \mn@doi [\apj] {10.3847/1538-4357/aae1f7}, 867, 43

\bibitem[\protect\citeauthoryear{Tokovinin \& Moe}{Tokovinin \&
  Moe}{2020}]{tokovinin_formation_2020}
Tokovinin A.,  Moe M.,  2020, \mn@doi [\mnras] {10.1093/mnras/stz3299}, 491,
  5158

\bibitem[\protect\citeauthoryear{{Truelove}, {Klein}, {McKee}, {Holliman},
  {Howell}  \& {Greenough}}{{Truelove}
  et~al.}{1997}]{truelove_resolution_1997ApJ...489L.179T}
{Truelove} J.~K.,  {Klein} R.~I.,  {McKee} C.~F.,  {Holliman} John~H. I.,
  {Howell} L.~H.,   {Greenough} J.~A.,  1997, \mn@doi [\apjl] {10.1086/310975},
  \href {https://ui.adsabs.harvard.edu/abs/1997ApJ...489L.179T} {489, L179}

\bibitem[\protect\citeauthoryear{{Welsh} et~al.,}{{Welsh}
  et~al.}{2012}]{welsh_kepler_2012Natur.481..475W}
{Welsh} W.~F.,  et~al., 2012, \mn@doi [\nat] {10.1038/nature10768}, \href
  {https://ui.adsabs.harvard.edu/abs/2012Natur.481..475W} {481, 475}

\bibitem[\protect\citeauthoryear{{Whitehouse} \& {Bate}}{{Whitehouse} \&
  {Bate}}{2006}]{whitehouse_2006MNRAS.367...32W}
{Whitehouse} S.~C.,  {Bate} M.~R.,  2006, \mn@doi [\mnras]
  {10.1111/j.1365-2966.2005.09950.x}, \href
  {https://ui.adsabs.harvard.edu/abs/2006MNRAS.367...32W} {367, 32}

\bibitem[\protect\citeauthoryear{{Whitehouse}, {Bate}  \&
  {Monaghan}}{{Whitehouse} et~al.}{2005}]{whitehouse_2005MNRAS.364.1367W}
{Whitehouse} S.~C.,  {Bate} M.~R.,   {Monaghan} J.~J.,  2005, \mn@doi [\mnras]
  {10.1111/j.1365-2966.2005.09683.x}, \href
  {https://ui.adsabs.harvard.edu/abs/2005MNRAS.364.1367W} {364, 1367}

\bibitem[\protect\citeauthoryear{{Whitworth}}{{Whitworth}}{1998}]{whitworth_jeans_instab_1998MNRAS.296..442W}
{Whitworth} A.~P.,  1998, \mn@doi [\mnras] {10.1046/j.1365-8711.1998.01479.x},
  \href {https://ui.adsabs.harvard.edu/abs/1998MNRAS.296..442W} {296, 442}

\bibitem[\protect\citeauthoryear{Wilson}{Wilson}{1927}]{wilson_1927}
Wilson E.~B.,  1927, \mn@doi [J. Am. Stat. Assoc.]
  {10.1080/01621459.1927.10502953}, 22, 209

\bibitem[\protect\citeauthoryear{Zagaria, Rosotti  \& Lodato}{Zagaria
  et~al.}{2021}]{zagaria_dust_2021}
Zagaria F.,  Rosotti G.~P.,   Lodato G.,  2021, \mn@doi [\mnras]
  {10.1093/mnras/stab985}, 504, 2235

\bibitem[\protect\citeauthoryear{Zanazzi \& Lai}{Zanazzi \&
  Lai}{2018}]{zanazzi_inclination_2018}
Zanazzi J.~J.,  Lai D.,  2018, \mn@doi [\mnras] {10.1093/mnras/stx2375}, 473,
  603

\bibitem[\protect\citeauthoryear{Zúñiga-Fernández
  et~al.,}{Zúñiga-Fernández et~al.}{2021}]{zuniga-fernandez_hd_2021}
Zúñiga-Fernández S.,  et~al., 2021, \mn@doi [\aap]
  {10.1051/0004-6361/202141985}, 655, A15

\bibitem[\protect\citeauthoryear{von Zeipel}{von
  Zeipel}{1910}]{von_zeipel_sur_1910}
von Zeipel H.,  1910, \mn@doi [Astronomische Nachrichten]
  {10.1002/asna.19091832202}, 183, 345

\makeatother
\end{thebibliography}


\appendix

\section{Notes on the mutual inclinations of observed systems}
\label{AppendixA}

In constructing Table \ref{tab:mutual_incs}, we have the following notes:

In the case of HD98800B, \citet{zuniga-fernandez_hd_2021} report two values for $i_\mathrm{disc}$ ($i_\mathrm{disc} = 26\degr$ or $154\degr$). They recover $\theta = 89 \pm 1 \degr$ when $i_\mathrm{disc} = 26\degr$, and $\theta = 134 \pm 1 \degr$ when $i_\mathrm{disc} = 154 \degr$. The alternate value of $\theta$ when $i_\mathrm{disc} = 26\degr$ is $\theta = 46 \pm 1 \degr$. In our statistics we use the higher value of $i_\mathrm{disc}$ as it recovers polar ($\theta = 91 \pm 3 \degr$) and retrograde values for $\theta$ whilst preserving our assumption of $180 \degr$ ambiguity in $\Omega_*$.

TWA 3A is complicated, there are many degenerate solutions as $i$ for both disc and inner binary are ambiguous. $\Omega$ for the disc is constrained, though the disc is either coplanar or polar depending on which $i$ you take. \citet{czekala_coplanar_2021} claim this disc to be coplanar, however they do not report the mutual inclination angle calculated with the degenerate values of $i$. When $i_* = 48.5 \pm 0.5 \degr$ and $i_\mathrm{disc}=131.2 \pm 0.7 \degr$, $\theta = 83.4 \pm 0.9 \degr$; when $i_* = 131.5 \pm 0.8 \degr$ and $i_\mathrm{disc} = 131.2 \pm 0.7 \degr$, $\theta = 9 \pm 6 \degr$. In the degenerative cases, the disc is either close to coplanar or in polar orbit. Assuming a $180 \degr$ ambiguity of $\Omega_*$ in these cases gives "alternate" values; in the coplanar case it is nearly polar, and in the polar case it is nearly retrograde. In this paper we use $i_* = 48.5$ and $i_\mathrm{disc} = 48.8$, and as with the other systems we give a "false" value of $\theta$ by assuming $180 \degr$ ambiguity in $\Omega_*$ even though this value has been constrained with RV data \citep{czekala_coplanar_2021}.

There is no value of $\Omega_*$ reported in the literature for R CrA \citep{mesa_exploring_2019,czekala_degree_2019} and so we do not include it in our statistics.

In the case of HD 142527 B, \citet{balmer_improved_2022} provide updated orbital parameters for this system. The corresponding mutual inclination angles are $\theta = 89.8 \pm 2 \degr$ and $\theta = 158.8 \pm 3 \degr$, so this disc is either polar or retrograde. This is not in agreement with the previous value of theta from \citet{czekala_degree_2019} of $\theta = 35 \pm 5 \degr$. We use the updated orbital parameters from \citet{balmer_improved_2022} in our statistics. The higher inclination is in better agreement with the modelling of \citet{price_circumbinary_not_transitional_2018MNRAS.477.1270P}.

\bsp	
\label{lastpage}
\end{document}